%% file: GPD_lecture_arXiv.tex
\documentclass[twocolumn]{svjour3}          
\usepackage[utf8]{inputenc}
\usepackage[T1]{fontenc}
\usepackage{amsmath}
\usepackage{amssymb}
\usepackage{xspace}
\usepackage[margin=2cm]{geometry}
\usepackage[numbers,sort&compress]{natbib}
\usepackage[svgnames]{xcolor}
\usepackage{graphicx}
\usepackage{changes}
\usepackage{slashed}
\usepackage{cuted}
\usepackage{etoolbox}
\usepackage{lipsum}

\newcommand{\ket}[1]{\ensuremath{\left|#1\right\rangle}\xspace}
\newcommand{\bra}[1]{\ensuremath{\left\langle #1\right|}\xspace}

\begin{document}

\title{An introductory lecture on Generalised Parton Distributions}



\author{Cédric Mezrag}


\institute{Cédric Mezrag \at
              Irfu, CEA, Université Paris-Saclay, F-91191, Gif-sur-Yvette, France \\
              \email{cedric.mezrag@cea.fr}
}

\date{Received: date / Accepted: date}

\maketitle

\begin{abstract}
These lecture notes on Generalised Parton Distributions aim at providing a general picture of the field on the theoretical and phenomenological sides to master and Ph.D. students. They go along with the lecture given at the Baryon International School of Physics in 2021. 
\end{abstract}

\section{Introduction : probing the internal structure of matter}
\label{sec:intro}

\input{section_introduction.tex}

\section{Generalised Partons Distributions}
\label{sec:GPDs}

\input{section_GPD.tex}

\section{Polynomiality and its consequences}
\label{sec:Polynomiality} 

\input{section_polynomiality.tex}

\section{Lightfront Wave Functions}
\label{sec:LFWFs}

\input{section_LFWFs.tex}

\section{Evolution}
\label{sec:Evolution}

\input{section_evolution.tex}

\section{Exclusive Processes}
\label{sec:Exclusive}

\input{section_exclusive.tex}

\section{Perspectives}
\label{sec:perspectives}

\input{section_perspectives.tex}

\begin{acknowledgements}
I am grateful to the Baryon International school of physics for the opportunity to give the lectures on GPDs, and also for their patience as this manuscript was delayed during its redaction. I would like to thanks V. Bertone, H. Dutrieux, J.M. Morgado Chavez, H. Moutarde, M. Riberdy and J.M. Rodriguez-Quintero for their valuable discussions and comments allowing me to improve the manuscript.  This project was supported by the European Union's Horizon 2020 research and innovation programme under grant agreement No 824093.  This work is supported in part in the framework of the GLUODYNAMICS project funded by the "P2IO LabEx (ANR-10-LABX-0038)" in the framework "Investissements d’Avenir" (ANR-11-IDEX-0003-01) managed by the Agence Nationale de la Recherche (ANR), France.
\end{acknowledgements}

\bibliographystyle{spphys}       
\bibliography{../../Bibliography}   

\end{document}

%% file: section_introduction.tex
Understanding the internal structure of objects is usually the most direct way to figure out the origin of the macroscopic properties. This is why many techniques have been developed through the centuries in order to produce images of physical or biological systems. 

\subsection{From macroscopic to microscopic structure}

A classical example of imaging for undergraduate students in physics is the Fraunhofer diffraction.
This optical phenomenon consists in observing the diffraction pattern in far field after diffracting usually visible and monochromatic light on an aperture or a grating.
The figure obtained is given by the Fourier transform of the aperture. 
The reader is probably familiar with the typical figures yielded by various apertures such as a slit, a double slit, a circular aperture, a grating, etc.
These examples can be found online and in multiple textbooks and are thus not reproduced here.
Nevertheless, we highlight that this classical imaging technique can be not only exploited with visible light but can also be generalised to smaller wavelengths.

Indeed, if visible light allows probing objects of sub-millimetre size, one can ask whether smaller systems are accessible by reducing the wavelength.
The answer is yes, as X-ray diffraction on crystals has allowed us to image the internal structure of these periodic systems.
The description of X-ray diffraction on crystals and the first experiments were performed by M. von Laue in Germany, and W. L. and W. H. Bragg in the UK (Nobel Prizes in 1914 and 1915 respectively).
The diffraction figure of these periodic systems presents spots directly related to the Fourier transform of the electronic density within the crystal, and thus to the geometry of the latter at the atomic scale. 

\subsection{Nucleon femtography}

\begin{figure}[b]
  \centering
  \includegraphics[width=0.5\textwidth]{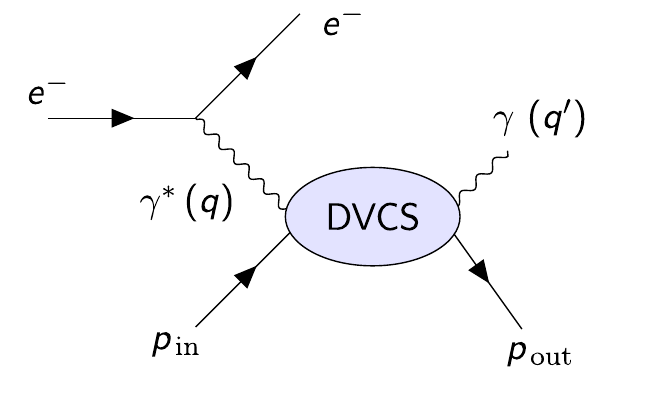}
  \caption{Deep Virtual Compton Scattering. The incoming proton interacts with the incoming electron through the exchange of a virtual photon. In the final state, the proton is left intact, and emits a real photon.}
  \label{fig:DVCS}
\end{figure}

From millimetre slit to atomic structure of crystals, one can realise that adapting the wavelength of the light diffracted on a system allows us to find back the geometry of the latter without breaking it.
Such a nice feature would deserve to be pursued at even smaller wavelength to probe subatomic matter, such as the partonic content of nucleons.
The immediate issue here is to provide a photon source of wavelength of a fermi or less.
Quantum field theory provides a solution through the use of virtual particles rather than real ones.
For instance, a high-energy electron beam can interact with a nucleon target by the exchange of a strongly virtual photon, \emph{i.e.} a photon whose momentum square $q^2$ is deeply spacelike $-q^2 = Q^2 \gg M_N^2$. This insures that the wavelength will be small enough to probe the internal structure of the nucleon. 
Then, providing that the struck nucleon is not broken and re-emits part of the incoming energy through a real photon, one ends-up with an experimental setup close to the classical diffraction systems.
This process, called Deep Virtual Compton Scattering (DVCS) \cite{Ji:1996nm,Radyushkin:1996nd}, is illustrated on figure \ref{fig:DVCS}.

Such a process is a good candidate to probe the internal structure of the nucleon.
However, the fact that the nucleon is a bound-state of quantum chromodynamics (QCD) strongly complicates the interpretation with respect to the much simpler cases of apertures or crystals mentioned before.
Indeed, quarks and gluons within the nucleon strongly interact with each other, and particles creations and annihilations happen continuously.
It is therefore unclear at this stage that any reliable pieces of information on the nucleon structure can be extracted through DVCS. 
In the following, we give a heuristic argument stating why DVCS can indeed pin down the partonic structure of the nucleon.

The difficulty of interpreting the DVCS in terms of a process probing the nucleon structure can be bypassed thanks to one of the more important properties of QCD: \emph{asymptotic freedom} \cite{Politzer:1973fx,Gross:1973id,Gross:1973ju}. Indeed, when the quark-gluon or gluon-gluon interactions happen at high energies, the strong coupling becomes small, allowing one to treat such an interaction within a perturbation framework. Since we work at a typical scale $Q^2\gg M_N^2$, we can indeed expect to use perturbation theory to describe the interaction between the deep virtual photon and the struck parton.

This struck parton receives therefore a large amount of momenta. In order to stay inside the proton and not be sent out and hadronise, parts of its energy has to be released. An easy possibility is to radiate this surplus of energy through the emission of a real photon. In this case, one really probes a single parton within the nucleon. But a second possibility would be to share this surplus of energy with other partons within the nucleon. It would trigger the exchange of quarks and gluons whose virtualities are large, of the order of $Q^2$. Such an involvement of multiple partons within the nucleon in a highly virtual interaction is possible, but less likely. In fact, it is suppressed\footnote{We do not consider here the Wilson line associated with the exchange of longitudinally polarised gluons.} in powers of $Q$. So for a large enough virtuality, the $\gamma^* p \to \gamma p$ amplitude can be decomposed as:

\begin{strip}
  \rule[-1ex]{\columnwidth}{1pt}\rule[-1ex]{1pt}{1.5ex}
  \begin{align}
    \label{eq:DVCSDecomp}
    \vcenter{\hbox{\includegraphics[width=0.2\textwidth]{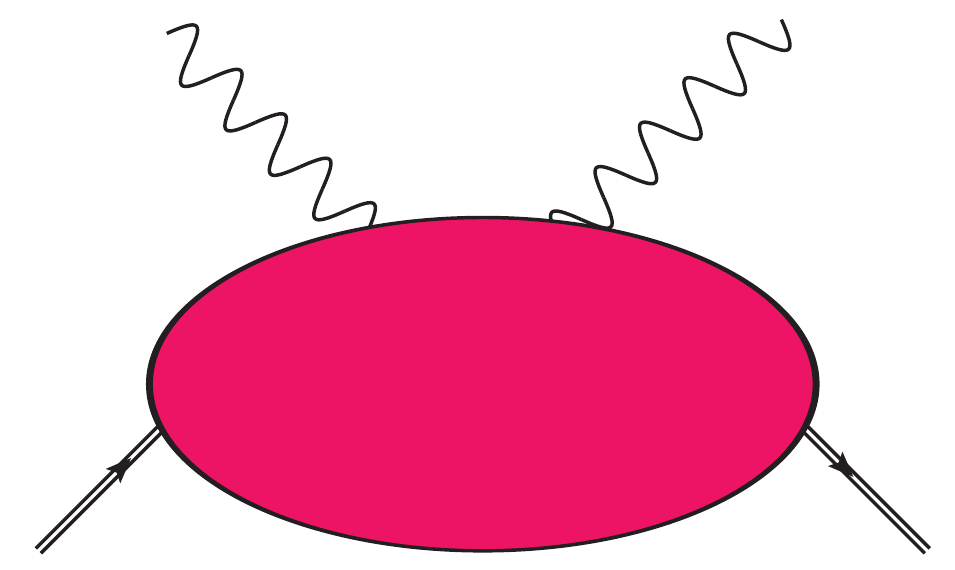}}}  = \vcenter{\hbox{\includegraphics[width=0.2\textwidth]{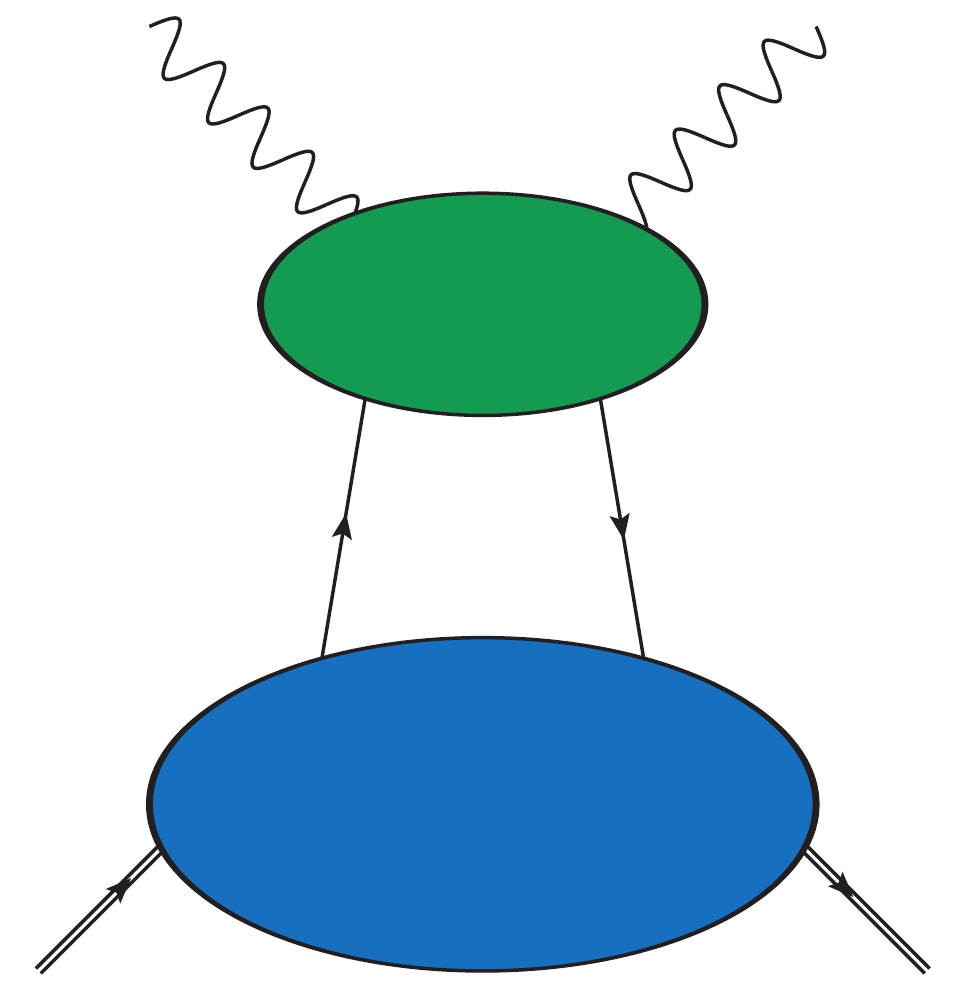}}}
    +\vcenter{\hbox{\includegraphics[width=0.2\textwidth]{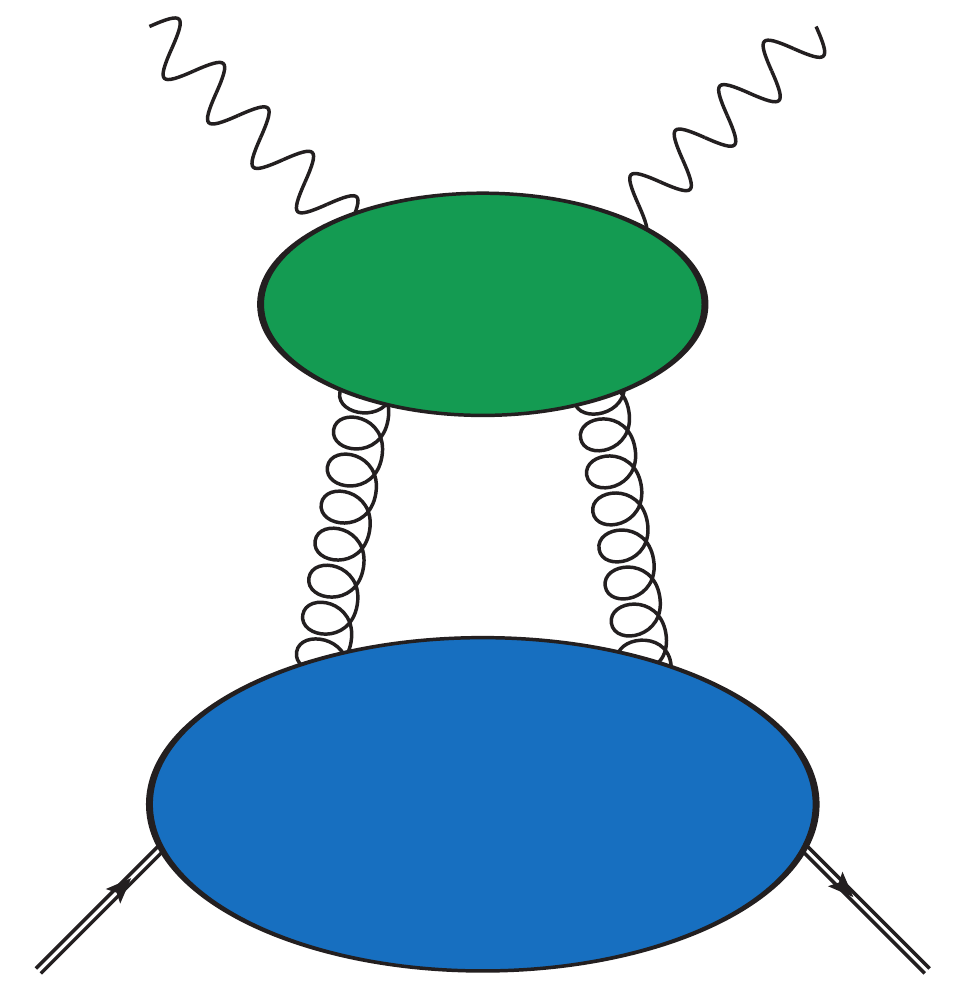}}}
    + \underbrace{\dots}_{\substack{\text{Power-suppressed} \\ \text{corrections}}} 
  \end{align}
  \hfill\rule[1ex]{1pt}{1.5ex}\rule[2.3ex]{\columnwidth}{1pt}
\end{strip}%
\noindent where the green ellipses represent the interaction between the singled out partons (quark or gluon) with the incoming virtual and outgoing real photon. It is where the high virtuality $Q^2$ circulates. Thus, it is computable in perturbation theory, higher-orders yielding logarithmic corrections in $Q^2$. Contrastingly, no physical hard scale is entering the description of the blue ellipses, which are thus not computable using perturbative QCD. They are instead parametrised using scalar, Lorentz invariant functions called Generalised Partons Distributions (GPDs) \cite{Mueller:1998fv,Ji:1996ek,Ji:1996nm,Radyushkin:1996ru,Radyushkin:1997ki}. As we will see below, among other very interesting properties, these functions can be interpreted in terms of multidimensional probability distributions in coordinate space through a Fourier transform, allowing finally to push below the femtometre scale the idea of mapping the internal structure of systems through light scattering.

In this heuristic description, we have avoided many technical difficulties justifying why we can really split DVCS between a hard part, computable in perturbation theory, and a non-perturbative contribution, parametrised in terms of GPDs. For instance, we have mentioned neither soft contributions and divergences nor longitudinally polarised gluon resummation in the Wilson line. The interested reader can find details in the literature \cite{Collins:1998be,Diehl:2003ny,Belitsky:2005qn}, as we will not develop these points further.


%% file: section_GPD.tex
In this section of the lecture, we will introduce Generalised Parton Distributions, present and motivate their main properties. We will focus on the case of the pion, which allows us to simplify the discussion, without removing substantial content regarding other hadrons. Reader interested in explicit formulae for other baryons are invited to consult the main references on GPDs such as \cite{Diehl:2003ny,Belitsky:2005qn}.

Throughout all the text, we will follow the notation convention introduced by M. Diehl \cite{Diehl:2003ny}. Our metric tensor $\eta^{\mu\nu} = \textrm{diag}(1,-1,-1,-1)$ and we define lightcone coordinates such that:
\begin{align}
  \label{eq:LightconeCoordinates}
  z^\pm & = \frac{1}{\sqrt{2}}(z^0\pm z^3), \quad z^\perp = (z^1, z^2),\\
  \label{eq:LightconeSquare}
  z^2 & = 2z^+ z^- - z_\perp^2
\end{align}

\subsection{Definitions and some properties}

GPDs have been introduced in the 1990s \cite{Mueller:1998fv,Ji:1996ek,Ji:1996nm,Radyushkin:1996ru,Radyushkin:1997ki}, as a ``parton-hadron'' $\to$ ``parton-hadron'' amplitude. The formal definition relies on a non-local matrix element, off-diagonal in momentum space, and where the quark or gluon fields composing the operator are separated by a light-like distance:
\begin{align}
  \label{eq:HqDef}
  H^q(x,\xi,t) = & \frac{1}{2} \int \frac{\textrm{d}z^-}{2\pi}e^{ixP^+z^-}\nonumber \\
                 & \times \bra{p_2}\bar{\psi}^q\left(-\frac{z^-}{2}\right) \gamma^+ \psi^q\left(\frac{z^-}{2} \right) \ket{p_1},\\
  \label{eq:HgDef}
  H^g(x,\xi,t)  = & \frac{1}{P^+} \int \frac{\textrm{d}z^-}{2\pi}e^{ixP^+z^-} \nonumber \\
                 & \times \bra{p_2}G^{+\mu}\left(-\frac{z^-}{2}\right) G^+_\mu\left(\frac{z^-}{2} \right) \ket{p_1},
\end{align}
where $\psi^q$ is a quark field of flavour $q$ and $G^{\mu\nu}$ is the gluon field strength. For convenience, we define the average momentum of the nucleon $P$ and the momentum transfer $\Delta$ as:
\begin{align}
  \label{eq:MomentumDef}
  P = \frac{p_1 + p_2}{2}, \quad \Delta = p_2-p_1.
\end{align}
%
%
The on-shell condition straightforwardly yields:
\begin{align}
  \label{eq:OnshellCondition}
  \left(P-\frac{\Delta}{2} \right)^2 =   \left(P+\frac{\Delta}{2} \right)^2 \Rightarrow P\cdot \Delta = 0, 
\end{align}
a relation which will be useful in the following.
%
From these definitions, we can provide an understanding of the kinematic variables $(x,\xi,t)$ on which the GPDs depend. Since $P$ is the average momentum carried by the incoming and outgoing pions, $x$, introduced in the argument of the exponential in eqs. \eqref{eq:HqDef} and \eqref{eq:HgDef} is the average momentum fraction along the lightcone of the active parton (\emph{i.e.} the one probed by the operator). The two other parameters, $(\xi,t)$ are defined from the momentum transfer $\Delta$ as:
\begin{align}
  \label{eq:DefXiandt}
  \xi = -\frac{\Delta^+}{2P^+}, \quad t = \Delta^2,
\end{align}
and therefore $t$ is the total momentum transfer (and corresponds to the standard Mandelstam variable in exclusive processes, hence the choice of notation), while $\xi$ encodes the deviation of the momentum fraction along the lightcone of the active parton with respect to the average momentum fraction.

\begin{figure}
  \includegraphics[height= 8cm, width= 8cm ]{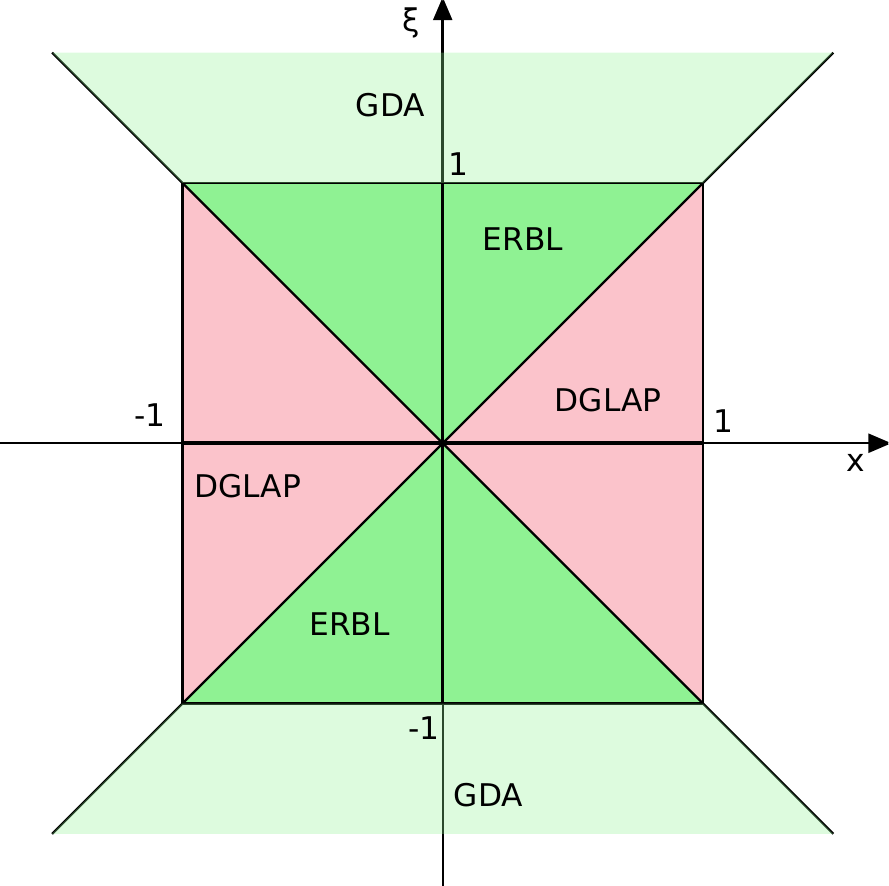}
\caption{GPD support in the $(x,\xi)$ plane. In pink the so-called DGLAP (or outer) region where $|x| \ge |\xi|$, and in green the ERBL (or inner) region where $|x| \le |\xi|$. The lighter green regions highlight the extensions of the ERBL region for $|\xi|\ge 1$ which correspond to the kinematic domain of GDA (up to the crossing symmetry).}
\label{fig:GPDPhaseSpace}       
\end{figure}
%
%

Because of its richer spin structure, more GPDs are necessary in order to parameterise the off-forward matrix element of the nucleon:
\begin{align}
  \label{eq:NucleonGPDquarks}
  & \frac{1}{2} \int \frac{\textrm{d}z^-}{2\pi}e^{ixP^+z^-}\bra{p_2}\bar{\psi}^q\left(-\frac{z^-}{2}\right) \gamma^+ \psi^q\left(\frac{z^-}{2} \right) \ket{p_1} \nonumber \\
  = & \frac{1}{2P^+}\bigg( H^q(x,\xi,t) \bar{u}(p_2) \gamma^+ u(p_1)  \nonumber \\
  & \left. \quad + E^q(x,\xi,t) u(p_2)\frac{i\sigma^{+\nu}\Delta_\nu}{2M}u(p_1)\right),
\end{align}
for quarks and
\begin{align}
  \label{eq:NucleonGPDgluons}
  & \frac{1}{P^+} \int \frac{\textrm{d}z^-}{2\pi}e^{ixP^+z^-}\bra{p_2}G^{+\mu}\left(-\frac{z^-}{2}\right) G^{+}_\mu\left(\frac{z^-}{2} \right) \ket{p_1} \nonumber \\
  = & \frac{1}{2P^+}\bigg( H^g(x,\xi,t)\bar{u}(p_2) \gamma^+ u(p_1)  \nonumber \\
  & \left. \quad + E^g(x,\xi,t) u(p_2)\frac{i\sigma^{+\nu}\Delta_\nu}{2M}u(p_1)\right).
\end{align}
Quark and gluon polarised distributions can also be defined for the nucleon (see \emph{e.g.} \cite{Diehl:2003ny}). However, since we do not use them in the following, we do not introduce them here.

%
%
Fig. \ref{fig:GPDPhaseSpace} may suggest that GPDs present a specific parity on variables $(x,\xi)$. Indeed, most of the GPDs are even in $\xi$ because of time reversal invariance (see \emph{e.g.} ref. \cite{Belitsky:2005qn} for a proof and list of exceptions for higher spin hadrons). Regarding the $x$-parity, gluons GPDs are even because gluons are their own antiparticles. However, quarks GPDs do not present any specific symmetries in $x$ and one defines:
\begin{align}
  \label{eq:NonSingletGPD}
  H^q_{NS} & = H^q(x,\xi,t) + H^q(-x,\xi,t),\\
  \label{eq:SingletGPD}
  H^q_{S} & =  H^q(x,\xi,t) - H^q(-x,\xi,t),
\end{align}
where $S$ stands for singlet combination and $NS$ for non-singlet one. These combinations play specific roles regarding the renormalisation properties of GPDs and are thus often introduced.
%
%
%

\begin{figure*}
  \begin{center}
    \includegraphics[width = 4cm, height= 3cm]{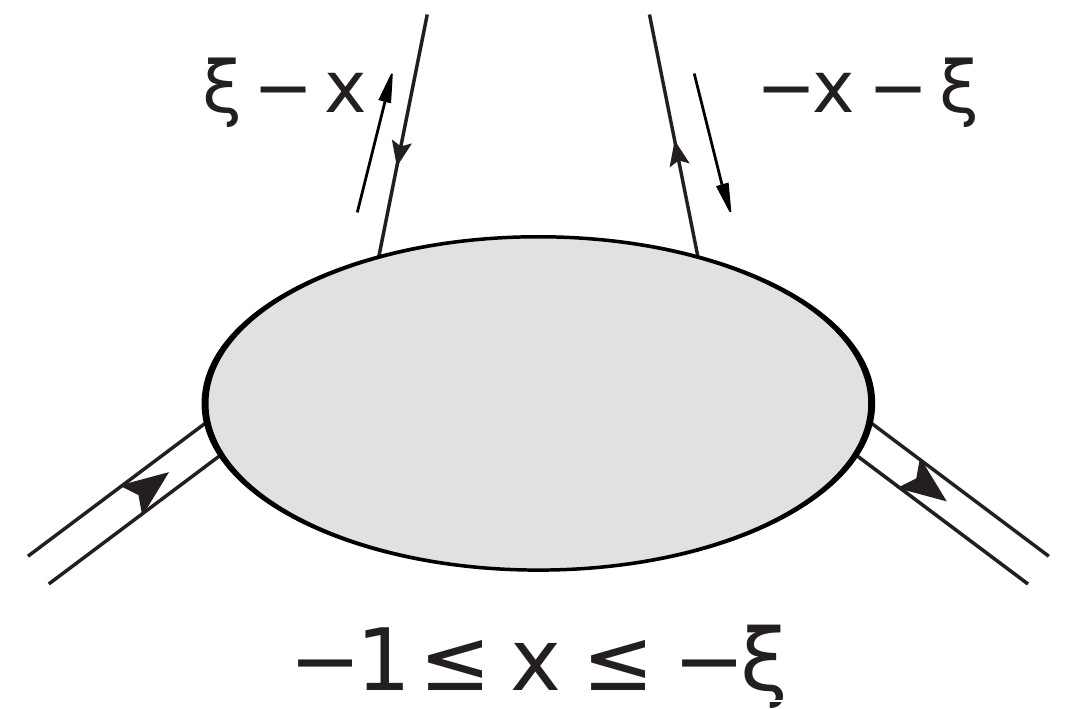}
    \includegraphics[width = 4cm, height= 3cm]{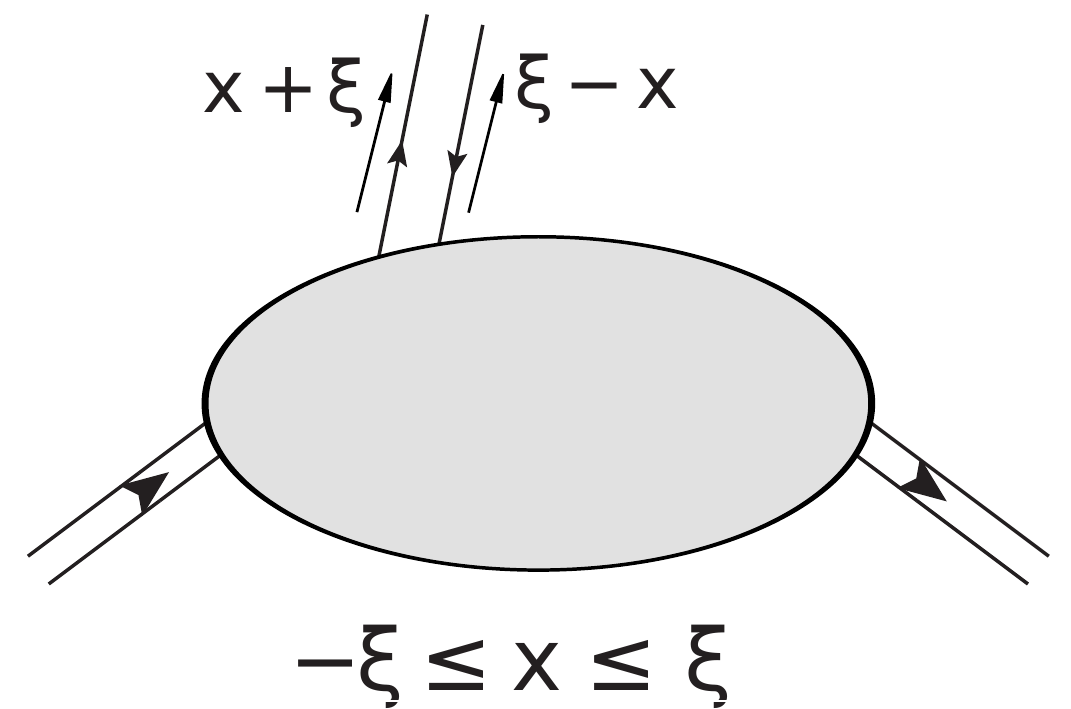}
    \includegraphics[width = 4cm, height= 3cm]{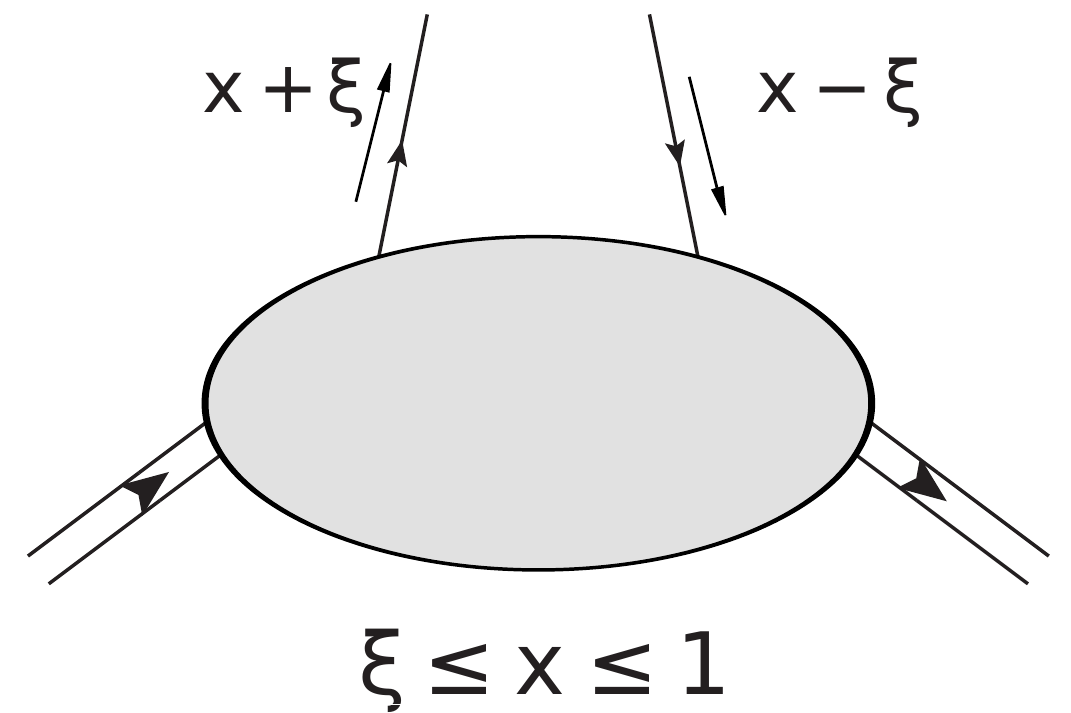}  
    \caption{Different interpretations of GPDs, depending on whether they are probed in the DGLAP or ERBL regions. From left to right, we first depict the case where $-1< x <- \xi$ so that the momentum fractions carried by the outgoing and incoming parton are both negative. It yields an antiquark interpretation of the parton probed. In the inner region $-\xi < x < \xi$, we represent the case where one momentum fraction is positive (quark interpretation), and one negative (antiquark interpretation), yielding the picture of the ERBL region as the ``extraction'' of a quark-antiquark pair from the hadron. On the right-hand, $\xi < x < 1$ and we find as expected the quark interpretation of the DGLAP region.  \label{fig:GPDLCInterpretation}}
  \end{center}
\end{figure*}

The fact that GPDs depend on two different lightcone momentum fractions $x$ and $\xi$ triggers interesting and unique consequences in the field of hadron physics.
Looking at $(x,\xi)$ plane, one can define several regions (see fig. \ref{fig:GPDPhaseSpace}).
In the entire plane, the GPD support is limited to the square $(x,\xi) \in [-1,1]^2$ as shown in ref. \cite{Diehl:1998sm}.
Within this square, we distinguish two types of regions: the outer one for which $|x| \ge |\xi|$, also known as the DGLAP (this acronym stands for Dokshitzer, Gribov, Lipatov, Altarelli, Parisi) and the inner one where $|\xi|\ge |x|$, known as the ERBL (Efremov-Radyushkin-Brodsky-Lepage) region.
These two regions carry different interpretations in terms of partons involved.
Indeed, focusing on the quark case, the outgoing and incoming active partons carry respectively a momentum along the lightcone of $(x \pm \xi ) P^+$. Depending on the sign of $(x \pm \xi)$ the incoming and outgoing partons will be seen either as a quark or an antiquark.
As illustrated on fig. \ref{fig:GPDLCInterpretation}, this yields three possibilities for $\xi>0$. If probed in the DGLAP region, our GPDs is seen as an active quark (or antiquark) being taken out and put back in the hadron. On the other hand, in the ERBL region, the GPD can be seen as a way to extract a pair of quark-antiquark from the hadron without breaking it.

These different interpretations have major consequences on the properties of GPDs, and in particular on their renormalisation properties. Anticipating on sec. \ref{sec:Evolution}, we can already highlight that the name of the two regions, DGLAP and ERBL, correspond to two different sets of evolution equations originally derived for Partons Distribution Functions (PDFs) \cite{Gribov:1972ri,Dokshitzer:1977sg,Altarelli:1977zs} and Distribution Amplitudes (DA) \cite{Efremov:1978rn,Efremov:1979qk,Lepage:1979zb,Lepage:1980fj} respectively. These evolution equations are found in the limit $\xi \to 0$ for DGLAP and $\xi \to 1$ for ERBL.

In addition, we mention that it is possible to extend formally the ERBL kinematic domain for $|\xi|\ge 1$. Up to crossing symmetry in the $s$ and $t$ Mandelstam variables, one obtains the Generalised Distribution Amplitudes (GDAs) \cite{Diehl:1998dk,Polyakov:1998ze,Kivel:1999sd,Diehl:2000uv} in that specific kinematic domain. The way to build this continuation will become clearer in sec. \ref{sec:Polynomiality}.

%
%

We will conclude this subsection by highlighting an important assumption we have made in our definitions in eq. \eqref{eq:HqDef} and \eqref{eq:HgDef}. Indeed, these definitions are valid in the so-called lightcone gauge, in which the $A^+$ component of the gluon field is set to zero\footnote{Technically, $A^+ = 0$ is not a sufficient condition to fully fix the gauge, and additional conditions need to be imposed (see \emph{e.g.} \cite{Hatta:2011ku}). }.
In other gauges, such as the covariant ones usually used when employing QCD perturbative and non perturbative techniques, one needs to add a Wilson line $\mathcal{W}$ defined as:
\begin{equation}
  \label{eq:WilsonLine}
  \mathcal{W}\left(-\frac{z^-}{2},\frac{z^-}{2}\right) = P \exp \left[ig \int_{-\frac{z^-}{2}}^{\frac{z^-}{2}} \textrm{d}\zeta^- A^+(\zeta^-) \right]
\end{equation}
where $P$ denotes the path ordering between $-z^-/2$ and $z^-/2$ and $g$ is the QCD coupling. The Wilson line ensures that the GPD definition is gauge invariant. In the following, we will keep working in the lightcone gauge, in which the Wilson line collapses to unity, and highlight differences where they appear in other gauges.

\subsection{Relation with PDFs, form factors and distribution amplitudes}

Looking closely at the definitions \eqref{eq:HqDef} and \eqref{eq:HgDef}, one realises that the operators are the same as the ones defining PDFs, the difference being only the non-vanishing momentum difference $\Delta$ between the incoming and outgoing hadron states. Taking $\Delta = 0$, the diagonal matrix element in momentum space is the one defining the PDFs. From eq. \eqref{eq:LightconeSquare} and \eqref{eq:DefXiandt}, one can see that this is equivalent to take the limit $(\xi,t)\to (0,0)$. Therefore, with our GPD definition, we obtain the following forward limits relating the pion GPDs with the quark $q$, antiquark $\bar{q}$ and gluon $g$ PDFs:
\begin{align}
  \label{eq:ForwardLimitQuark}
  H^q(x,0,0) & = q(x) \Theta(x) - \bar{q}(-x) \Theta(-x), \\
  \label{eq:ForwardLimitGluon}
  H^g(x,0,0) & = x g(x) \Theta(x) - x g(-x) \Theta(-x),
\end{align}
where $\Theta$ is the Heaviside function. This forward limit has been at the core of GPD modelling strategies in the past decades \cite{Vanderhaeghen:1999xj, Musatov:1999xp, Goloskokov:2005sd, Mezrag:2013mya, Chavez:2021llq, Raya:2021zrz}, as unpolarised PDFs are nowadays well known from phenomenological extractions on a wide kinematic range in $x$.

Another interesting property of GPDs is their connection with the electromagnetic form factors (EFFs). Indeed, when considering the matrix element defining the GPDs in eq. \eqref{eq:HqDef}, one realises that the specific case $z^-=0$ corresponds to the distance along the lightfront between the quarks fields vanishing and the operator becoming \emph{local}. Because of the Fourier transform, such a configuration can be selected through a simple integration over $x$:
\begin{align}
  \label{eq:EFF}
  \int \textrm{d}x\, & H_\pi^q(x,\xi,t) \nonumber \\
  & = \int \! \frac{\mathrm{d}z^-\delta(P^+z^-)}{2}\bra{p_2}\bar{\psi}^q(-\frac{z}{2})\gamma^+\psi^q(\frac{z}{2})\ket{p_1}\nonumber \\
                                      & = \frac{1}{2P^+} \bra{p_2}\bar{\psi}^q(0)\gamma^+\psi^q(0)\ket{p_1}.
\end{align}
This constraint has also been heavily used in GPD phenomenological modelling, as it allows us to connect elastic scattering data with the $t$-dependence of GPD.

The two previous constraints apply for all hadrons, with differences depending on their spin. The following property is specific to the pion. Indeed, one can relate the Distribution Amplitude (DA) $\varphi$ \cite{Efremov:1978rn,Efremov:1979qk,Lepage:1979zb,Lepage:1980fj} of the pion with the pion quark GPD in the limit $(\xi,t)\to (1,0)$. The original proof, introduced in ref. \cite{Polyakov:1998ze} relies on GDAs rather than GPDs, and exploits the partial conservation of axial current (PCAC). This allows us to relate GDAs to standard DAs, and through crossing symmetry, to extend this property to GPDs. This property is known as the GPD soft pion theorem and impacts the singlet and non-singlet combinations such that:
\begin{align}
  \label{eq:NonSingletSoftPion}
  H^q_{NS}(x,1,0) & = \varphi\left(\frac{1+x}{2} \right), \\
  \label{eq:SingletSoftPion}
  H^q_{S}(x,1,0) & = 0.
\end{align}
Anticipating a bit on sec. \ref{sec:Evolution}, a direct consequence of eq. \eqref{eq:SingletSoftPion} is that the gluon GPD is also vanishing in the limit $(\xi,t)\to (1,0)$, contrasting again with all other hadrons. Finally, let us mention that an alternative proof was providing in ref. \cite{Mezrag:2014jka}, without relying on GDAs, but on a specific truncation of the pion Bethe-Salpeter kernel. 

\subsection{Interpretation in coordinate space}

Until now, we have restricted ourselves to work in momentum space, the reason being that experimental amplitudes related to GPDs are computed within momentum space. But it is also possible to define GPDs in the so-called impact parameter space (see the works of M. Burkardt \cite{Burkardt:2000za} and M. Diehl \cite{Diehl:2002he}), \emph{i.e.} when the Fourier transform of the momentum component in the transverse plane is taken. It yields a mixed representation: $x$ is a degree of freedom in momentum space, while $b_\perp$ is introduced and represent the position in the transverse plane. Being bound states, hadrons present a spatial extension, and thus we need to define what we mean by its position in the transverse plane. A lightcone surrogate for the centre of mass in multi-body classical systems is required. 

Without entering too much in technical details (which can be found in ref. \cite{Diehl:2002he}), let us mention that we can build an analogy between transverse boost (\emph{i.e.} boost along vector living in the transverse plane) and 2D Galilean transformations. It allows us to define a centre of ``plus-momentum'' $b_\perp$ in the transverse plane as a weighted average of the position $b_{i;\perp}$ off all partons composing the hadron of interest. We thus obtain:
\begin{align}
  \label{eq:CenterofPlusMomentum}
  b_{\perp} = \frac{\sum_i k_i^+ b_{i;\perp}}{\sum_i k_i^+}
\end{align}
where the $k_i$ are the momenta of the partons within the hadron. Through this procedure, we can define a ``centre of plus-momentum'' for both the incoming $b_\perp^{in}$ and outgoing $b_\perp^{out}$ hadron. Because of the incoming hadron carrying a momentum $(1+\xi)P^+$ along the lightcone and the outgoing one $(1-\xi)P^+$, one expects $b_\perp^{in} \neq b_{\perp}^{out}$ and concludes that GPDs are also off-diagonal in impact-parameter space. 
\begin{figure}[t]
  \centering
  \includegraphics[height=7cm,width = 7cm]{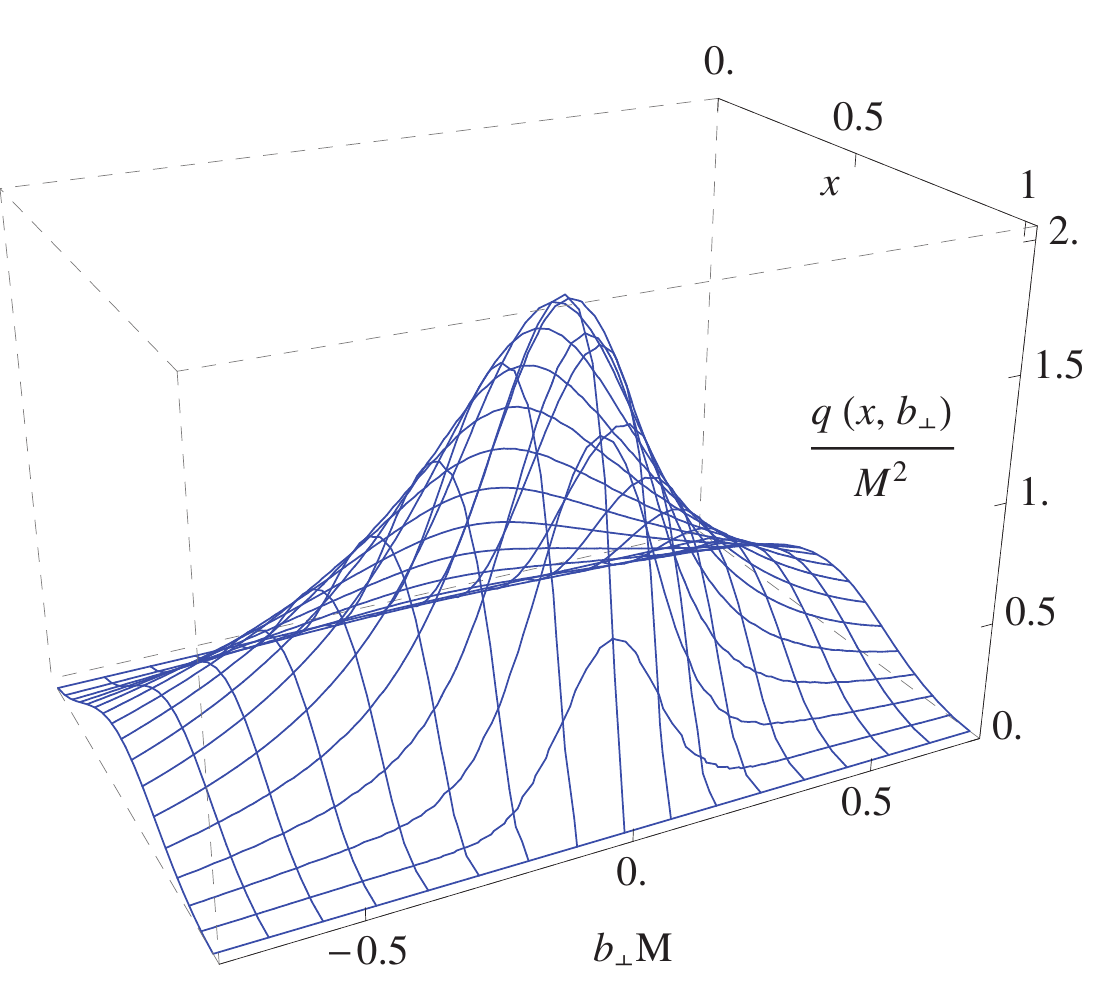}
  \caption{Pion GPD interpretation in the impact parameter space. Figure from \cite{Mezrag:2014jka}.}
  \label{fig:Pion3DPicture}
\end{figure}

Since $b_\perp^{in} \neq b_{\perp}^{out}$, we define a vector $\bar{b}_\perp$ as the Fourier conjugate of:
\begin{equation}
  \label{eq:DperpDef}
  D_\perp = \frac{p_{2\perp}}{1-\xi}-\frac{p_{1\perp}}{1+\xi},
\end{equation}
introduced such as $t$ depends on $p_{1\perp}$ and $p_{2\perp}$ only through $D_\perp$\cite{Diehl:2002he}. Using this new variable, one can express $b_\perp^{in}$ and $b_{\perp}^{out}$ as:
\begin{align}
  b_\perp^{in} =  \frac{\xi\bar{b}_\perp}{1+\xi}, \quad b_{\perp}^{out} = -\frac{\xi \bar{b}_\perp}{1-\xi}, 
\end{align}
and the operators defining the GPDs are evaluated at a position $\bar{b}_\perp$ in the transverse plane, \emph{i.e.} a parton is created and annihilated at position $\bar{b}_\perp$. Within this notation, we see that the centre of plus momentum coincide between the two hadron states when $\xi = 0$. This has some major consequences, as in that case, GPDs become diagonal in impact parameter space, and one obtains a density interpretation in 1 dimension in momentum space ($x$) and 2 dimensions in coordinate space $b_\perp$ \cite{Burkardt:2000za,Diehl:2002he}. For the pion, this density $q(x,\bar{b}_\perp)$ is given as:
\begin{align}
      q(x,\bar{b}_\perp) = \int \frac{\textrm{d}^2\Delta_\perp}{(2\pi)^2} e^{i \Delta_\perp \bar{b}_\perp}H(x,0,-\Delta_\perp^2).
\end{align}
Fig.  \ref{fig:Pion3DPicture} illustrates the density that can be found in model computations (see ref.  \cite{Mezrag:2014jka} for details). Experimental extractions have also been attempted (see for instance ref. \cite{Moutarde:2018kwr} for a recent example on the nucleon). However, these extractions suffer three main difficulties:
\begin{itemize}
\item the interpretation of exclusive processes in terms of GPDs through collinear factorisation is valid only at values of $t$ such that $-t << Q^2$ where $Q$ is the hard scale of the considered process;  
\item however, one needs to integrate $t$ up to infinity to perform the Fourier transform, introducing model-dependent extrapolations;
\item experimental data are accessible only at non vanishing $\xi$, requiring therefore an extrapolation toward $\xi =0$, which again generates model-dependence and biases.
\end{itemize}
Notwithstanding these difficulties, the ability to perform multidimensional tomography of the nucleon through GPDs remains strongly appealing. Consequently, GPDs are parts of the core physics case of current and future experimental facilities, such the US \cite{AbdulKhalek:2021gbh} and Chinese \cite{Anderle:2021wcy} electron ion colliders. 

\subsection{Connection with the Energy-Momentum Tensor}

In addition to delivering a unique opportunity to perform a 2+1D tomography of the nucleon, GPDs present another strongly valuable feature: they are connected to the hadrons Energy-Momentum Tensor (EMT). The latter, labelled $T$, is defined and parameterised for the pion as \cite{Polyakov:2018zvc}:
\begin{align}
  \label{eq:EMTDefPion}
  \bra{p_2}T^{\mu\nu}_{a}\ket{p_1}& =  2P^\mu P^\nu A^{a}(t) \nonumber \\
                                 & +\frac{1}{2} \left(\Delta^\mu\Delta^\nu-\eta^{\mu\nu}\Delta^2\right)C^{a}(t) \nonumber \\
                                 &+ 2 M^2 \eta^{\mu\nu}\bar{C}^{a}(t)
\end{align}
where the form factors $A$, $C$ and $\bar{C}$ are sometimes called gravitational form factors (although the space is \emph{flat} and thus no gravitational effect is taken into account), the index $a$ indicates quark $q$ or gluon $g$ contributions and $M$ is the hadron mass.

In the nucleon case, additional terms enter the definition (see \emph{e.g.} \cite{Bakker:2004ib,Leader:2013jra}):\hfill
\begin{strip}
  \rule[-1ex]{\columnwidth}{1pt}\rule[-1ex]{1pt}{1.5ex}
  \begin{align}
    \label{eq:EMTdefProton}
    \bra{p_2} T^{\mu\nu}_a(0)\ket{p_1}
    = \bar u(p_2) \Bigg\{&
                                   \frac{P^\mu P^\nu}{M}\,A^a(t)
                                   + \frac{\Delta^\mu\Delta^\nu - \eta^{\mu\nu}\Delta^2}{M}\, C^a(t)
                                   + M \eta^{\mu\nu}\bar C^a(t) \nonumber
    \\
                                 &+ \frac{i\left(P^{\mu} \sigma^{\nu\rho}+P^{\nu} \sigma^{\mu\rho}\right)\Delta_\rho}{4M}\left[A^a(t)+B^a(t)\right]+ \frac{P^{[\mu} i\sigma^{\nu]\rho}\Delta_\rho}{4M}\,D^a(t)\Bigg\} u(p_1) \, ,
  \end{align}
  \hfill\rule[1ex]{1pt}{1.5ex}\rule[2.3ex]{\columnwidth}{1pt}
\end{strip}%
where $a^{[\mu}b^{\nu]}=a^\mu b^\nu-a^\nu b^\mu$. 
These form factors have to obey specific constraints coming from conservation laws \cite{Ji:1996ek,Brodsky:2000ii,Lowdon:2017idv,Lorce:2019sbq}:
\begin{align}
  \label{eq:ConservationLawA}
  &\sum_f A^{q_f}(0) +A^g(0) = 1 \\
    \label{eq:ConservationLawB}
  &\sum_f B^{q_f}(0) +B^g(0) = 0 \\
    \label{eq:ConservationLawCb}
  &\sum_f \bar{C}^{q_f}(t) +\bar{C}^g(t) = 0
\end{align}
where $\sum_f$ is the sum over the considered quark flavours. The connection to GPDs is provided by the computation of the the following Mellin moment:
\begin{align}
  \label{eq:MomentEMTHquark}
  \int_{-1}^1 \textrm{d}x\, x H^{q}(x,\xi,t) = A^q(t) + 4\xi^2 C^q(t),
\end{align}
connecting two out of three EMT form factors to GPDs in the case of the pion. We will justify the polynomial dependence in $\xi$ of this formula in the sec. \ref{sec:Polynomiality}. For the pion, only the $\bar{C}$ form factor remains inaccessible through leading-twist GPDs. Regarding the nucleon, an additional sum rule holds in the quark sector:
\begin{align}
\label{eq:MomentEMTEquark}
  \int_{-1}^1 \textrm{d}x\, x E^{q}(x,\xi,t) = B^q(t) - 4\xi^2 C^q(t)
\end{align}
allowing us to obtain three out of five EMT form factors for the nucleon\footnote{Polarised GPDs provide an additional constraint on the EMT form factor $D$, see \emph{e.g.} refs. \cite{Lorce:2017wkb,Dutrieux:2021nlz} so that again, only $\bar{C}$ remains unconstrained by GPDs.}. In the gluons sector, a similar relation can be written between GPDs and EMT form factors:
\begin{align}
    \label{eq:MomentEMTHgluon}
  \int_{-1}^1 \textrm{d}x\, H^{g}(x,\xi,t) = A^g(t) + 4\xi^2 C^g(t),
\end{align}
for both the pion and the nucleon, and
\begin{align}
\label{eq:MomentEMTEgluon}
  \int_{-1}^1 \textrm{d}x\, E^{g}(x,\xi,t) = B^g(t) - 4\xi^2 C^g(t)
\end{align}
for the nucleon. 

These sum rules have a major consequence for the nucleon. Indeed, one can show that the total angular momentum carried by each quark flavour $J^q$ and gluons $ J^g$ is given by \cite{Ji:1996ek}:
\begin{align}
  \label{eq:Jquarks}
  J^q & = \frac{1}{2}\left( A^q(0) + B^q(0)\right), \\
  \label{eq:Jgluons}
  J^g & = \frac{1}{2}\left( A^g(0) + B^g(0)\right).
\end{align}
This yields the so-called Ji sum rules allowing to connect quark and gluon GPDs to their contribution to the total angular momentum of the nucleon:
\begin{align}
  \label{eq:JiSumRulequarks}
  2 J^q & = \int_{-1}^1\textrm{d}x\, x \left(H^q(x,\xi,0)+E^q(x,\xi,0) \right), \\
  \label{eq:JiSumRulegluons}
  2 J^g & = \int_{-1}^1\textrm{d}x\, \left(H^g(x,\xi,0)+E^g(x,\xi,0) \right),
\end{align} 
where the contribution of the $C$ form factors cancels.  

If the $C$ and $\bar{C}$ EMT form factors do not contribute to the angular momentum of the nucleon, they carry an interesting interpretation in terms of pressure and shear forces within the nucleon \cite{Polyakov:2002yz,Polyakov:2018zvc,Lorce:2018egm,Dutrieux:2021nlz}. Following \cite{Dutrieux:2021nlz}, we focus here on the isotropic pressure $p_q$ and $p_g$ generated by quarks and gluon respectively, and the pressure anisotropy $s_q$ and $s_g$ due to quarks and gluons. In terms of $C$ and $\bar{C}$ form factors, they can be expressed in the Breit frame (\emph{i.e.} the frame where the system is on average at rest) for the nucleon as \cite{Dutrieux:2021nlz}:
\begin{align}
  \label{eq:IsotropicPressure}
  p_i(\vec{r}) & = M \int \frac{\textrm{d}^3 \vec{\Delta}}{(2\pi)^3}e^{-i \vec{r} \vec{\Delta} }\left[\frac{2t}{3M}C^i(t) - \bar{C}^i(t) \right],\\
  \label{eq:AnisotropyPressure}
  s_i(\vec{r}) & = \frac{4M}{r^2} \int \frac{\textrm{d}^3 \vec{\Delta}}{(2\pi)^3} e^{-i \vec{r} \vec{\Delta} }\frac{t^{-1/2}}{M^2}\frac{\textrm{d}^2}{\textrm{d}t^2}\left[t^{5/2} C^i(t)\right] ,
\end{align}
where $\vec{r}$ is a spatial three-vector and the index $i$ stand for quark flavours or gluons. Similar relations can be derived for the pion \cite{Polyakov:2018zvc}. From eqs. \eqref{eq:IsotropicPressure} and \eqref{eq:AnisotropyPressure}, the $s_i$ are independent of the $\bar{C}$ form factor and can thus be entirely characterised by GPDs. Moreover the total pressure distribution within the nucleon (\emph{i.e.} summed over quark and gluon contributions) is also independent of $\bar{C}$ because of eq. \eqref{eq:ConservationLawCb} and can also be in principle extracted from GPDs. 

Before concluding this section, let us highlight that in the chiral limit, the soft pion theorem  given in eq. \eqref{eq:SingletSoftPion} allows us to constrain the $C$ EMT form factor of the pion at vanishing momentum transfer $t$ as:
\begin{align}
  \label{eq:SoftPionFFSinglet}
   0 = \int_{-1}^1 \textrm{d}x\, x H_S^{q}(x,1,0) = A^q(0) + 4 C^q(0).
\end{align}
Moreover, because of evolution equations mixing singlet and gluon contributions (see sec.\ref{sec:Evolution}), $H_S^q$ will remain zero only if the gluon GPD is itself vanishing. Assuming that the soft pion theorem is scale independent, the constrain in eq. \eqref{eq:SoftPionFFSinglet} can therefore be generalised to the gluon case.


%% file: section_polynomiality.tex
In the previous section, we have encountered two types of sum rules connecting the Mellin moment of GPDs to the EFFs in eq. \eqref{eq:EFF} and to the EMT form factors in eqs. \eqref{eq:MomentEMTHquark} to \eqref{eq:MomentEMTEgluon}. In this section, we will generalised the results to higher Mellin moments and highlight the underlying mathematical structure and its consequences. 

\subsection{Local operators}

We define the $m$-th Mellin moment $\mathcal{M}_m$ of the GPD as:
\begin{equation}
  \label{eq:MellinMomentDef}
  \mathcal{M}_m(\xi,t) = \int_{-1}^1 \textrm{d}x \, x^m H(x,\xi,t). 
\end{equation}
Not only the $0^{th}$ one corresponding to the EFFs can be directly connected to local operators, but such a connection can be generalised for all $m$: 
\begin{align*}
  & \mathcal{M}_m(\xi,t) \nonumber \\
  = &  \frac{1}{2}\int \frac{ \textrm{d}x\mathrm{d}z^- \, x^m  e^{ixP^+z^-}}{2\pi}\bra{p_2}\bar{\psi}^q(-\frac{z^-}{2})\gamma^+\psi^q(\frac{z^-}{2})\ket{p_1}\nonumber \\
  = & \int \frac{\textrm{d}x\, \textrm{d}z^-}{2(iP^+)^m} \frac{\textrm{d}^m}{(\textrm{d}z^-)^m}\left[\frac{e^{ixP^+z^-}}{2\pi }\right] \nonumber \\
  & \quad \quad  \times \bra{p_2}\bar{\psi}^q(-\frac{z^-}{2})\gamma^+\psi^q(\frac{z^-}{2})\ket{p_1} \nonumber \\
  = & \frac{i^m}{2(P^+)^{m+1}} \left. \bra{p_2}\frac{\textrm{d}^m}{(\textrm{d}z^-)^m}\left[\bar{\psi}^q(-\frac{z^-}{2})\gamma^+\psi^q(\frac{z^-}{2})\right]\ket{p_1}\right|_{z=0}\nonumber \\
  = & \frac{1}{2(P^+)^{m+1}} \bra{p_2}\bar{\psi}^q(0)\gamma^+\left(i \overleftrightarrow{\partial}^+ \right)^m\psi^q(0)\ket{p_1},
\end{align*}
where
\begin{equation}
  \label{eq:partialleftright}
  \overleftrightarrow{\partial}^\mu = \frac{1}{2}\left(\overrightarrow{\partial}^\mu - \overleftarrow{\partial}^\mu \right).
\end{equation}
In the case the gauge chosen is not the lightcone one, then the partial derivative $\overleftrightarrow{\partial}^\mu$ is replaced by the covariant derivative $\overleftrightarrow{D}^\mu$ where $D^\mu = \partial^\mu -igA^\mu$. This modification is a consequence of the Wilson line.

Introducing a lightlike vector $n$ such that $n \cdot a  = a^+$ and $n^2 = 0$, one can rewritten the Mellin moments as: 
\begin{align}
  \label{eq:refMMoperator}
 \mathcal{M}_m & = \frac{1}{ 2(P^+)^{m+1}} n_\mu n_{\mu_1}\dots n_{\mu_m} \nonumber \\
               & \quad \times \bra{p_2}\bar{\psi}^q(0)\gamma^\mu i \overleftrightarrow{\partial}^{\mu_1}\dots i \overleftrightarrow{\partial}^{\mu_m}\psi^q(0)\ket{p_1}, \\
  \label{eq:refMMoperator2}
               & = \frac{1}{ 2(P^+)^{m+1}} n_\mu n_{\mu_1}\dots n_{\mu_m} \nonumber \\
               & \quad \times \bra{p_2}\bar{\psi}^q(0)\gamma^{\{\mu} i \overleftrightarrow{\partial}^{\mu_1}\dots i \overleftrightarrow{\partial}^{\mu_m\}}\psi^q(0)\ket{p_1},
\end{align}
where $a^{\{\dots\}}$ stand for a symmetrised and traceless operator over the Lorentz indices.
One can indeed see that permuting two Lorentz indices in the operator of eq. \eqref{eq:refMMoperator} does not modify the contraction.
In the same way, tracing the operator with $\eta_{\mu_i\mu_j}\eta^{\mu_i\mu_j}$ yields a contribution proportional to $n^2$ and thus vanishing. 

Looking at eq. \eqref{eq:refMMoperator2}, one can recognise the twist-two local operators $O^{\mu \mu_1\dots \mu_m}$ introduced in the description of Deep Inelastic Scattering:
\begin{align}
  \label{eq:DefLocalOperators}
  O^{\mu \mu_1\dots \mu_m} = \bar{\psi}^q(0)\gamma^{\{\mu} i \overleftrightarrow{\partial}^{\mu_1}\dots i \overleftrightarrow{\partial}^{\mu_m\}}\psi^q(0),
\end{align}
but this time evaluated between off-diagonal hadron states in momentum space.
This is of course a major difference with PDFs as the parametrisation of the off-forward local matrix elements for the pion is given as:
\begin{align}
  \label{eq:ParaMellinMoments}
  \bra{p_2}&\bar{\psi}^q(0)\gamma^{\{\mu} i \overleftrightarrow{\partial}^{\mu_1}\dots i \overleftrightarrow{\partial}^{\mu_m\}}\psi^q(0)\ket{p_1} \nonumber \\
           & = P^{\{\mu}\sum_{i=1}^{m+1} P^{\mu_1}\dots P^{\mu_{i-1}} \Delta^{\mu_i}\dots \Delta^{\mu_m\}} A^q_{i,m}(t) \nonumber \\
           & \quad + \Delta^\mu \Delta^{\mu_1}\dots \Delta^{\mu_m} C^q_{m+1}(t).
\end{align}
Several comments can be made regarding this parametrisation.
First the Lorentz structure is built from the two available degrees of freedoms, $P$ and $\Delta$, combined so that the symmetry and traceless properties of the operator are maintained. This Lorentz structure is multiplied by Lorentz Scalar $A_{i,m}$ and $C_{m+1}$ called generalised form factors. Like the EFFs, these form factors also depend on Lorentz scalar, the one available being $P^2$, $\Delta^2$ and $P\cdot \Delta$. From eq. \eqref{eq:OnshellCondition} we know that $P\cdot \Delta = 0$, and thus that $P^2$ and $\Delta^2$ are not independent but obey the on-shell condition $P^2+ \Delta^2/4 = M^2$. Therefore, like the EFFs, the generalised form factors depend only on $t$.  Moreover, applying discrete symmetries, one can furthermore constrain the generalised form factors. Since GPDs are even in $\xi$, then only the tensorial structures containing even powers of $\xi$ survive.

One can thus write the Mellin moments of GPDs as:
\begin{align}
  \label{eq:Polynomiality}
  \mathcal{M}_m (\xi,t) = \sum_{i=0}^{\left[\frac{m}{2}\right]}& (2\xi)^{2i}A^q_{i,m}(t) \nonumber \\
  & + \textrm{mod}(m,2) (2\xi)^{m+1}C_{m+1}(t),
\end{align}
where $[\dots ]$ is the floor function and $\textrm{mod}(2,m)$ is 1 if $m$ is odd, zero otherwise. Eq. \eqref{eq:Polynomiality} yields the so-called polynomiality property \cite{Ji:1998pc,Radyushkin:1998bz}, stating that the Mellin moments of the GPDs are polynomials in $\xi$.

\subsection{The $D$-term and the Radon transform}

In eq. \eqref{eq:Polynomiality}, we have, on purpose, attributed a specific label $C$ to the generalised form factors coming with the highest power in $\xi$.
Indeed, this contribution to the highest power in $\xi$ deserves to be highlighted among all the generalised form factors.
First, it vanishes for even powers of $m$. This means that it is generated solely by the singlet combination of the GPDs (see eq. \eqref{eq:SingletGPD}), which is odd. In fact, we can define a generating odd function $D$ called the $D$-term\footnote{The $D$-term $D(y,t)$ should not be confused with the EMT form factor $D(t)$. In fact, as we will see in the following sections, the $D$-term is connected with the EMT for factor $C$. This discrepancy in the notation is unfortunate but common in the literature, we therefore present it as such in these lecture notes.} such that:
\begin{align}
  \label{eq:DtermDef}
        \int_{-1}^1 \textrm{d}y\, y^m D(y,t) = (2)^{m+1}C_{m+1}(t).
\end{align}
At this level, we \emph{assume} that such a function exists and is unique\footnote{The delicate question of existence and uniqueness are intricated. In mathematics, we face here a problem related to the Hausdorff moment problem whose solution exists and is unique providing that the series of moments is completely monotonic \cite{Hausdorff:1921a,Hausdorff:1921b}. However our situation is more involved, as $D$ is not positive definite. We therefore assume the existence and uniqueness of $D$.}. We can then write the polynomiality property of eq. \eqref{eq:Polynomiality} as:
\begin{align}
  \label{eq:PolynomialityDterm}
  \sum_{i=0}^{\left[\frac{m}{2}\right]} (2\xi)^{2i}A^q_{i,m}(t) = &\int_{-1}^1 \textrm{d}x\, x^mH(x,\xi,t)\nonumber \\
                                                                  & \quad -\xi^{m+1}\int_{-1}^1 \textrm{d}y\, y^m D(y,t).
\end{align}
Simply changing the variable $y\to x/\xi$ and assuming that $\xi > 0$, one can write the previous equation as:
\begin{align}
  \label{eq:LugwigHegalson}
  \sum_{i=0}^{\left[\frac{m}{2}\right]} (2\xi)^{2i}A^q_{i,m}(t) = & \int_{-1}^1 \textrm{d}x\, x^m \bigg[H(x,\xi,t) \nonumber \\
  & \quad \quad \left.- D\left(\frac{x}{\xi},t\right)\mathbb{I}_{-\xi\le x \le \xi } \right],
\end{align}
where $\mathbb{I}_{-\xi\le x \le \xi }$ is 1 for $x\in [-\xi;\xi]$ and 0 otherwise. Eq. \eqref{eq:LugwigHegalson} has four \emph{major} consequences. Firstly, it tells us that the $D$-term is not a general function of $x$ and $\xi$ but a function of the ratio $x/\xi$. Secondly, the $D$-term lives only in the ERBL region, and vanishes in the DGLAP ones. It presents therefore a symmetric role with respect to the PDF, as it obeys the ERBL evolution equations when the PDFs follow the DGLAP ones. Moreover, looking at eqs. \eqref{eq:MomentEMTHquark} to \eqref{eq:MomentEMTEgluon}, the first moment of the $D$-term is directly proportional to the EMT form factor $C$ (hence the notation $C_{m+1}$ in eq. \eqref{eq:ParaMellinMoments}). It connects the $D$-term with basic properties of hadrons, and highlights the fact that such a function should not be ignored in attempts to model GPDs. 

Finally, let us highlight the most important outcome of eq. \eqref{eq:LugwigHegalson}. The power of the polynomial in $\xi$ in the left-hand side, is the same as the order of the moment on the right-hand side. This situation is known in mathematics as the Lugwig-Hegalson consistency condition. Hertle showed in the 1980 \cite{Hertle:1983} that this is a necessary and sufficient condition so that $H-D$ is in the range of the Radon transform $\mathcal{R}$ \cite{Radon:1917,Radon:1917tr}. In other words, it exists a function $F$ such that $\mathcal{R}[F] = H-D$.
Introduced independently by D. Mueller \cite{Mueller:1998fv} and A. Radyushkin \cite{Radyushkin:1997ki}, $F$ is called a Double Distribution (DD).
Its was identified as a Radon amplitude in ref. \cite{Teryaev:2001qm}. The Radon transform formalism for GPDs was further developed in ref. \cite{Chouika:2017dhe}. The explicit connection between $H-D$ and $F$ is thus given as:
\begin{align}
  \label{eq:RadonTransform}
  & H(x,\xi,t) - \mathbb{I}_{-\xi\le x \le \xi } D\left(\frac{x}{\xi},t \right) \nonumber \\
  = & \int_\Omega \textrm{d}\beta \textrm{d}\alpha \, \delta(x-\beta - \alpha \xi) F(\beta,\alpha,t)
\end{align}
where $\Omega = \{(\beta,\alpha)\, | \, |\alpha|+ |\beta| \le 1\}$. One can then reabsorb the $D$-term on the right-hand side yielding the well-known formula connecting DD to GPDs:
\begin{align}
  \label{eq:DDtoGPD}
  H(x,\xi,t) = \int_\Omega  &\textrm{d}\beta \textrm{d}\alpha \, \delta(x-\beta - \alpha \xi)\nonumber \\
  & \times \left[F(\beta,\alpha,t) +\xi \delta(\beta) D(\alpha,t)\right].
\end{align}
Up to minor modifications, the same results hold for gluon GPDs, and thus one can define a gluon $D$-term and gluon Double Distributions. Regarding the nucleon, an additional DD needs to be added to describe the GPD $E$, but the \emph{same} $D$-term appear in the Mellin Moment of $H$ and $E$, up to a minus sign. The situation is therefore similar to the pion case. 

Last but not least, let us add that the $D$-term was originally introduced in the context of double distributions \cite{Polyakov:1999gs}, as a possible complementary tensorial structure to the DD $F$.

\subsection{Double Distributions}

The double distribution $F$ was originally introduced as a way to parameterise a matrix element \cite{Radyushkin:1997ki}. For the pion it yields:
\begin{align}
  \label{eq:DDDef}
  &\left. \bra{p_2}\bar{\psi}(\frac{-z}{2})\gamma \cdot z \psi(\frac{z}{2})\ket{p_1}\right|_{z^2 = 0} \nonumber \\
  = & 2 P\cdot z \int \textrm{d}\beta \textrm{d}\alpha e^{-i\beta P\cdot z + i \alpha \Delta \cdot z /2} F(\beta,\alpha,t) \nonumber \\
  & - \Delta\cdot z   \int \textrm{d}\beta \textrm{d}\alpha e^{i \alpha \Delta \cdot z /2}D(\alpha,t)
\end{align}
where $D$ was added subsequently in \cite{Polyakov:1999gs} as mentioned above. One can then check that using the definition of DD in eq. \eqref{eq:DDDef} and choosing $z$ such that only the $z^-$ is not vanishing, the Fourier transform over $z^-$ defining the GPD in \eqref{eq:HqDef} give back the Radon transform in eq. \eqref{eq:DDtoGPD}.

\begin{figure}[t]
  \centering
  \includegraphics[width=8cm, height=8cm]{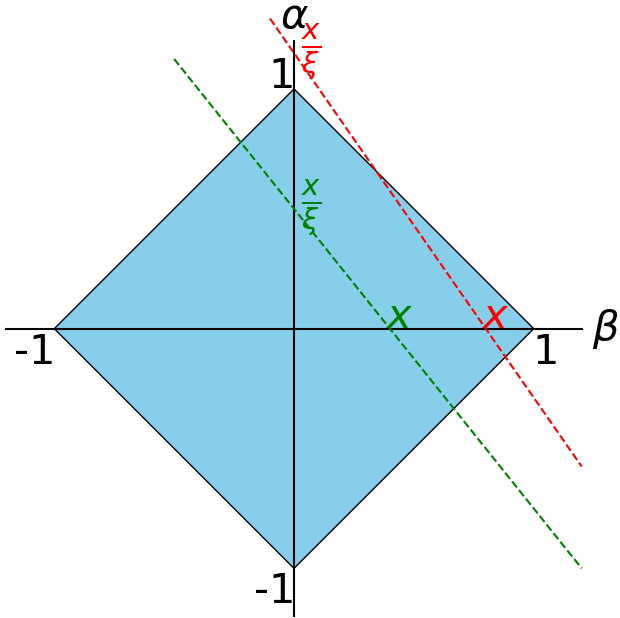}
  \caption{Support of the Double Distribution $F$ in the $(\beta,\alpha)$ space. In red, an integration line corresponding to a DGLAP region kinematic; in green an integration line corresponding to a ERBL kinematic.}
  \label{fig:DDsupport}
\end{figure}

The properties of GPDs are naturally extended to the DDs. The even parity in $\xi$ is associated with an even parity in $\alpha$ for $F$ and an odd parity for $D$ that we already highlighted. The support property in $(\beta,\alpha)$ space can also be derived from the one of the GDPs in $(x,\xi)$ space. Guaranteeing that the GPDs is zero in the DGLAP region for $x >1 $ imposes that the DD support is a rotated square in the $(\beta,\alpha)$ space labelled $\Omega$, and illustrated on fig. \ref{fig:DDsupport}.

From that point, we can draw a geometrical interpretation for the connection of the DGLAP and ERBL region in the $(\beta,\alpha)$ space. From eq. \eqref{eq:DDtoGPD}, we see that the connection between DDs and GPDs is made through integration along a line $\beta = x-\alpha\xi$. Therefore, lines whose slopes are steeper than 1, crossing the $\beta=0$ axis outside the support region contribute solely to the GPDs DGLAP region (red line on fig. \ref{fig:DDsupport}). On the other hand, lines whose slope are softer than 1 cross the $\beta = 0$ axis within the support region and contribute to the ERBL region in $(x,\xi)$ space (green line on fig. \ref{fig:DDsupport}). As the reader has probably already noticed, the lines intersect the $\alpha=0$ axis at the point $\beta = x$. Following our geometrical interpretation, we directly see that DGLAP lines intersecting the $\alpha=0$ axis at values of $\beta >1$ will yield a vanishing contribution as they never cross the DD support, in agreement with the GPD support property. However, ERBL-type lines present a non-vanishing contribution, even for $|x| >1$. These lines correspond also to $|\xi|\ge 1$ which make them contribute to the GDA region, already pointed out in fig. \ref{fig:GPDPhaseSpace}. Double Distributions therefore present a natural way to combine both GPDs and GDA in a single description. 

Other properties can be directly generalised from GPDs to DDs. For instance, the GPDs forward limit in terms of PDF can be directly understood at the level of DDs. Setting $\Delta = 0$ in eq. \eqref{eq:DDDef} or in \eqref{eq:DDtoGPD}, one realises that the DD $F$ and the PDF $q$ are connected through:
\begin{equation}
  \label{eq:DDForwardLimit}
  q(x) = \int^{1-|x|}_{-1+|x|} \textrm{d} \alpha F(x,\alpha,0).
\end{equation}
This property has been extensively used in modelling attempts of DDs (see sec. \ref{sec:DDModels}).

In addition to the PDFs, it is interesting to highlight the connection of the DDs with the local operators and the generalised form factors. Inserting the relation between GPDs and DDs of eq. \eqref{eq:DDDef} into eq. \eqref{eq:MellinMomentDef}, one gets:

\begin{strip}
  \rule[-1ex]{\columnwidth}{1pt}\rule[-1ex]{1pt}{1.5ex}
    \begin{align}
      \label{eq:MMrevisited}
      \int_{-1}^1 \textrm{d}x \, x^m H(x,\xi,t) = & \int_\Omega \textrm{d}\beta \textrm{d}\alpha \, (\beta+\alpha \xi)^m F(\beta,\alpha,t) + \xi^{m+1}\int_{-1}^1 \textrm{d}\alpha \alpha^m D(\alpha,t)\nonumber \\
      = & \sum_{i=0}^{[\frac{m}{2}]} \xi^{2i} \underbrace{\binom{m}{2i} \int_\Omega \textrm{d}\beta \textrm{d}\alpha\, \alpha^{2i}\beta^{m-2i}F(\beta,\alpha,t)}_{= (2)^{2i} A_{i,m}(t)} \quad + \xi^{m+1}\underbrace{\int_{-1}^1 \textrm{d}\alpha \alpha^m D(\alpha,t)}_{= (2)^{m+1}C_{m+1}(t)}.
    \end{align}
  \hfill\rule[1ex]{1pt}{1.5ex}\rule[2.3ex]{\columnwidth}{1pt}
\end{strip}%
\noindent Eq. \eqref{eq:MMrevisited} provides us with an interesting interpretation of the generalised form factors. They are the bidimensional Mellin moments of the Double Distribution $F$, which can now be seen as the generating function of the set of generalised form factors.

From eq. \eqref{eq:DDForwardLimit} and \eqref{eq:MMrevisited}, we see that we can also write the Mellin moments as:
\begin{align}
  \label{eq:GaugeMM}
  \mathcal{M}(\xi,t)  = & A_{0,m}(t) \nonumber \\
  &  + \xi \left[\sum_{i=1}^{[\frac{m}{2}]}(2\xi)^{2i}A_{i,m}(t) +2(2\xi)^m C_{m+1}(t)\right],
\end{align}
where $A_{0,m}(t)$ are the Mellin moments over $x$ of the $t$-dependent PDF $q(x,t)$. Applying the same procedure than before, we can prove that it exists a function $G$ such as $(H-q)/\xi$ is the Radon transform of $G$ and thus:
\begin{align}
  \label{eq:DefDDG}
  H(x,\xi,t)  = &q(x,t)+ \xi \int_\Omega \textrm{d}\beta \textrm{d}\alpha G(\beta,\alpha,t)\delta(x-\beta-\alpha \xi)) \nonumber \\
              =  & \int_\Omega \textrm{d}\beta \textrm{d}\alpha\delta(x-\beta-\alpha \xi)) \nonumber \\
  & \quad \times \left[\delta(\alpha)q(\beta,t)  + \xi G(\beta,\alpha,t)\right].
\end{align}
This has two main consequences: i) comparing the results with eq. \eqref{eq:DDtoGPD}, it shows that the way to define DDs is not unique and that we have previously made a specific choice by singling the $D$-term out of the DD; and ii) a striking similarity between the PDF and the $D$-term can be highlighted, both being the one-variable reduction of the DD $F$ and $G$:
\begin{align}
  \label{eq:PDFReduction}
  q(x,t) & = \int_{-1+|x|}^{1-|x|}\textrm{d}\alpha F(\beta,\alpha,t), \\
    \label{eq:DtermReduction}
  D(x,t) & = \int_{-1+|x|}^{1-|x|}\textrm{d}\beta G(\beta,\alpha,t).
\end{align}
In fact we face here the DD \emph{scheme} ambiguity\footnote{Also called gauge ambiguity in the literature}, which was firstly highlighted in \cite{Teryaev:2001qm}, and then studied in the literature \cite{Tiburzi:2004qr,Chouika:2017dhe}. To fully describe the GPDs, two DD, $F$ and $G$, are needed. However, most of the generalised form factors can be described either through $F$ or $G$. Only the $t$-dependent PDF and the $D$-term are intrinsically associated to $F$ and $G$ respectively and cannot be ``exchanged'' in a scheme transformation such as those highlighted in \cite{Teryaev:2001qm,Tiburzi:2004qr,Chouika:2017dhe}. It highlights once again the specificity of both the $q(x,t)$ and $D(x,t)$.

\subsection{Models of Double Distributions}
\label{sec:DDModels}

Double Distributions are a popular way to model GPDs because, as we saw above, they are equivalent to fulfilling the polynomiality property. In this section, we present classical and modern ways to model DDs.

\subsubsection{The Radyushkin Double Distribution Ansatz}

The most popular technique to model DDs is certainly the Radyushkin Double Distribution Ansatz (RDDA). Introduced in the early days of GPDs phenomenology \cite{Musatov:1999xp}, it consists in choosing the scheme in which $G$ is reduced to the $D$-term, following eq. \eqref{eq:DDtoGPD}, and modelling $F$ through a multiplicative Ansatz:
\begin{equation}
  \label{eq:RDDA}
  F(\beta,\alpha,t) = q(\beta,t) \pi_N(\beta,\alpha),
\end{equation}
where $\pi_N$ is the so-called profile function depending on a parameter $N$ and responsible for generating the skewness dependence of the model. It takes the functional form:
\begin{equation}
  \label{eq:ProfileFunction}
  \pi_N(\beta,\alpha) = \frac{\Gamma\left(N+\frac{3}{2}\right)}{\sqrt{\pi}\Gamma(N+1)}\frac{((1-|\beta|)^2-\alpha^2)^N}{(1-|\beta|)^{2N+1}},
\end{equation}
and is normalised such that:
\begin{equation}
  \label{eq:ProfileFunctionNormalisation}
   \int_{-1+|\beta|}^{1-|\beta|}\textrm{d}\alpha \, \pi_N(\beta,\alpha) = 1,
\end{equation}
guaranteeing therefore the sum rule eq. \eqref{eq:PDFReduction}. Different modelling assumptions have been suggested for the $t$-dependent PDF, from the simple product of the PDF with the EFF \cite{Vanderhaeghen:1999xj}, to Regge-motivated phenomenological models \cite{Goeke:2001tz}. It was used in the case of the pion \cite{Amrath:2008vx,Chavez:2021llq,Chavez:2021koz} to provide phenomenologically-motivated pion GPDs models and also in the case of the nucleon, where it was used to extract, in a model-dependent way, GPDs from exclusive processes \cite{Guidal:2008ie,Goloskokov:2007nt}. The success of this Ansatz comes from the facts that it is simple to implement, that it is driven by the PDF and EFFs which are much better known quantities, and that it generates phenomenologically meaningful results. However, for fit purposes, the flexibility of the Ansatz was somehow limited. Consequently, attempts to exploit it in other DD schemes have been performed and yielded significantly different GPDs \cite{Radyushkin:2011dh,Mezrag:2013mya,Radyushkin:2013hca}.  

\subsubsection{Other types of models}

\begin{figure}[t]
  \centering
  \includegraphics[width=5cm,height=6cm]{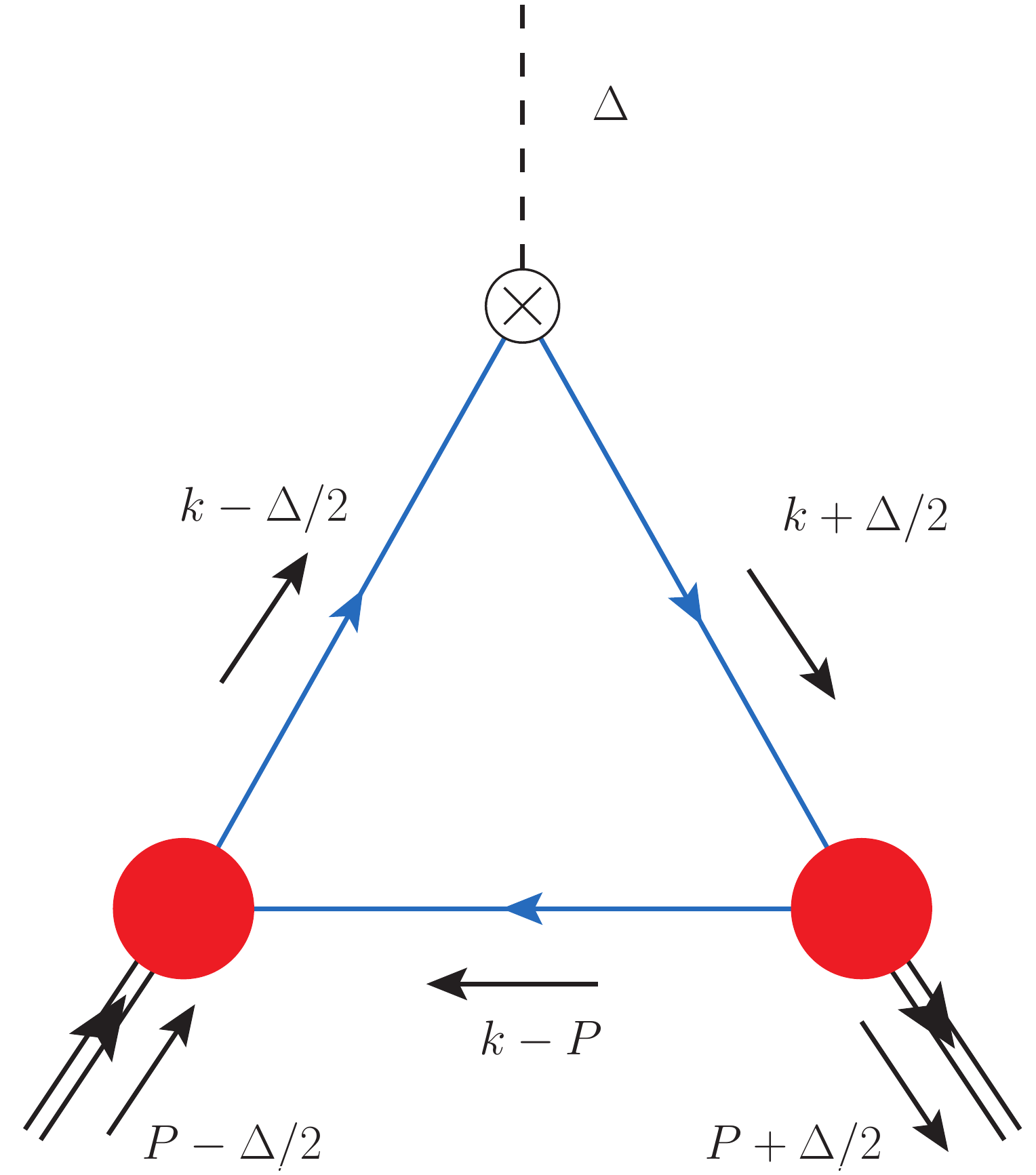}
  \caption{Typical diagram found in attempts to compute the pion GPD in a diagrammatic approach. When covariantly treated, these diagrams exhibit the expected Radon transform structure.}
  \label{fig:TriangleDiagrams}
\end{figure}

Double Distributions appear in different types of computations. We will here simply mention the diagrammatic modelling attempts, which consist in computing Feynman-like diagrams in which the hadron is decomposed into partons, one of whom being the active one (see the example for the pion on fig. \ref{fig:TriangleDiagrams}). When using covariant propagators and effective vertices, the Lorentz structure is preserved, and thus the computations exhibit the features of the Radon transform as expected \cite{Tiburzi:2002tq,Tiburzi:2004mh,Chang:2014lva,Mezrag:2014tva,Mezrag:2014jka,Mezrag:2016hnp}.

Within these studies, the Double Distributions exhibited can be significantly more complicated than the RDDA, and appear in schemes in which neither $F$ nor $G$ are reduced to solely the $t$-dependent PDF or the D-term \cite{Mezrag:2014tva,Mezrag:2016hnp}. Other diagrammatic approaches have been attempted, especially using the Nambu-Jona-Lasinio model, but without explicitly exhibiting the DDs \cite{Broniowski:2007si,Freese:2019eww,Freese:2020mcx}.


%% file: section_LFWFs.tex
The DD modelling strategy provides an interesting way to obtain GPDs fulfilling by construction the polynomiality property. But other modelling strategies exist, with their own advantages and drawbacks. In this section, we present the one relying on the Light-front wave functions (LFWFs), as new techniques based on covariant bound-state equations have been developed to compute them \cite{Carbonell:2017kqa, Mezrag:2017znp, Mezrag:2018hkk, Zhang:2021mtn, Raya:2021zrz, Eichmann:2021vnj, dePaula:2022pcb}.

\subsection{Lightfront Wave Function decomposition of GPDs}

The starting point of the LFWFs modelling is the decomposition of hadrons states in terms of these LFWFs, labelled here $\Phi_\beta^{i\dots j}$ such that:
\begin{align}
  \label{eq:FockDecMeson}
  \ket{P,\pi} &\propto \sum_\beta \Phi_\beta^{q\bar{q}}\ket{q\bar{q}}+\sum_\beta \Phi_\beta^{q\bar{q},q\bar{q}}\ket{q\bar{q},q\bar{q}} + \dots \\
  \label{eq:FockDecBaryon}
  \ket{P,N} &\propto \sum_\beta \Phi_\beta^{qqq}\ket{qqq}+\sum_\beta \Phi_\beta^{qqq,q\bar{q}}\ket{qqq,q\bar{q}} + \dots 
\end{align}
where $\beta$ is a generic label for the relevant quantum numbers and $^{i\dots j}$ denotes the partonic content of the state associated with the LFWF considered. Within such an expansion, the entire non-perturbative physics of a given quantum fluctuation corresponding to a possible Fock state is contained within the associated LFWF. A LFWF of $N$ partons, $\Phi^N$ depends then on $N$ momentum fraction along the lightcone $\{x_1,\dots, x_N\}$ such that their sum is equal to one, but also on $2N$ momentum variables in the transverse plane, labelled $\{k_\perp^1,\dots k_\perp^N\}$, where $k_\perp$ is a 2-vector. By momentum conservation, one has $P_\perp = \sum_{i}^N k^i_\perp$.

The matrix element defining GPDs and given in eq. \eqref{eq:HqDef}, \eqref{eq:HgDef}, \eqref{eq:NucleonGPDquarks} and \eqref{eq:NucleonGPDgluons} can then be expressed in terms of Fock states and LFWFs. Since the derivation is given in great details in ref. \cite{Diehl:2000xz} in the general case and in related work for a truncated Fock space extension (see for instance refs. \cite{Mezrag:2016hnp,Chouika:2017rzs}), we will not reproduce it here.

\begin{figure}[t]
  \centering
  \includegraphics[width=0.45\textwidth]{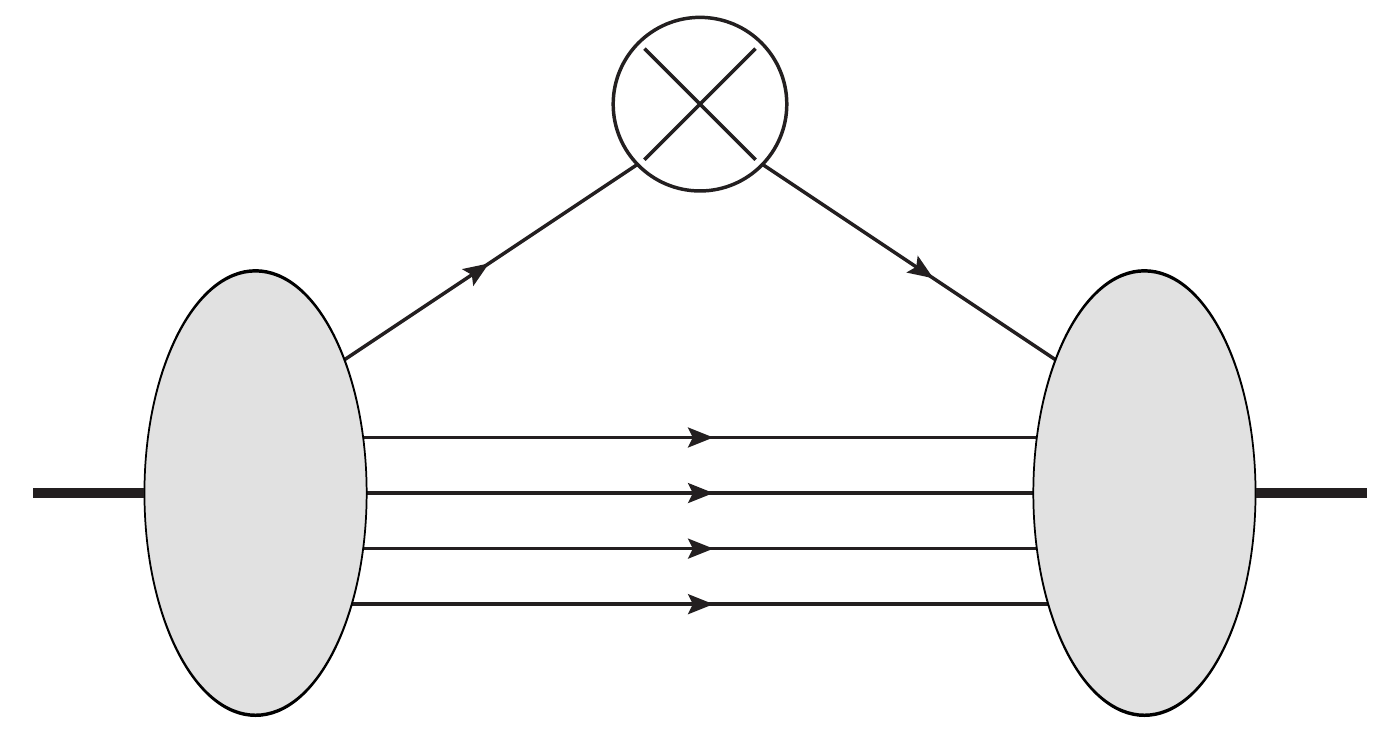}\\
  \includegraphics[width=0.45\textwidth]{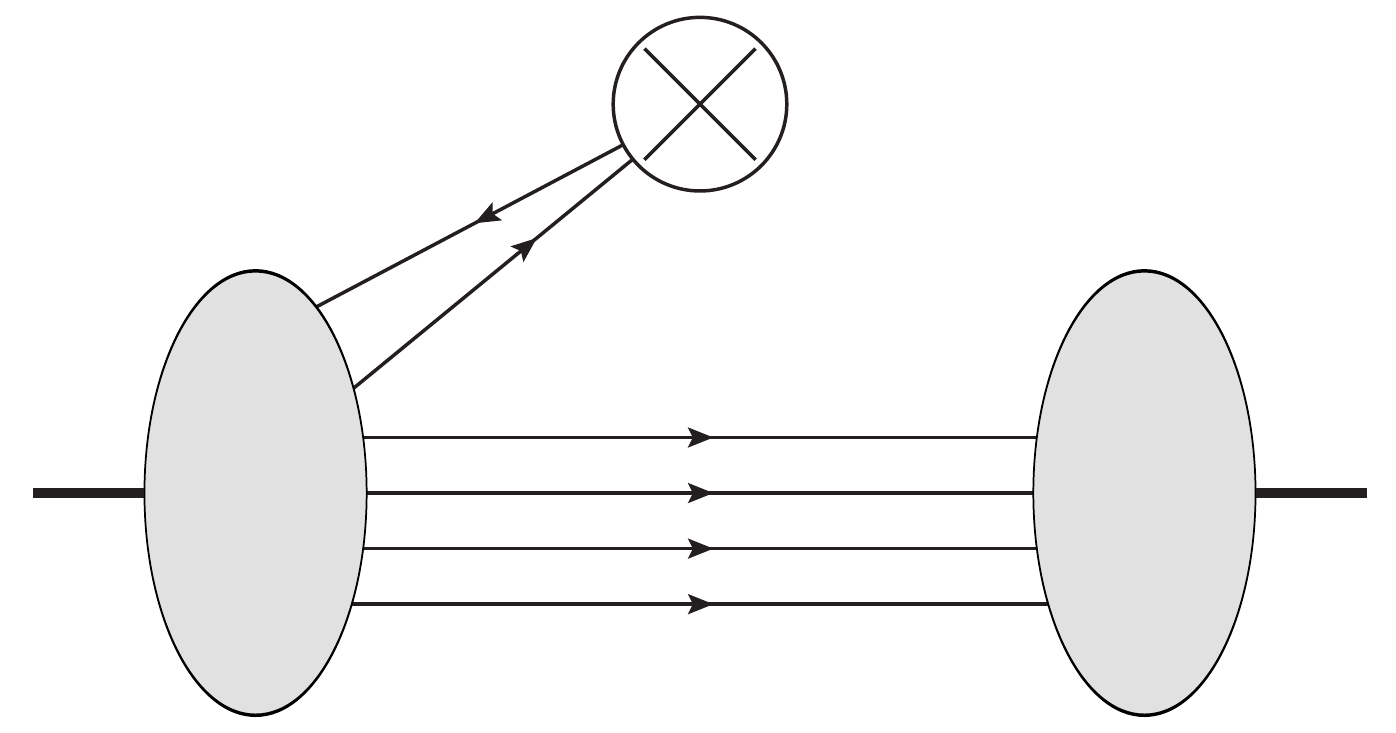}
  \caption{LFWFs decomposition of GPDs. Upper panel, DGLAP region interpretation conserving parton number between incoming and outgoing states; lower panel, ERBL region interpretation not conserving parton number as the incoming state emits a pair of quark-antiquark.}
  \label{fig:OverlapPicture}
\end{figure}

Instead, we will highlight the fact that since the interpretation of the GPDs in terms of partonic content depends on the kinematic area (see fig. \ref{fig:GPDLCInterpretation}), the physical content of the overlap is also modified by these different interpretations. In the DGLAP region, the overlap representation yields a LFWFs representation that we can sketch as:
\begin{align}
  \label{eq:DGLAPOverlap}
  \left.H^q(x,\xi)\right|_{x\ge \xi} \propto & \sum_N \sqrt{1-\xi^2}^{1-N} \sum_j \delta_{s_j,q}\int \left[\textrm{d}x_i \textrm{d}^2k_\perp^i \right]_N \nonumber \\
  & \times\left(\Phi^N (\hat{r}_N)\right)^* \Phi^N(\tilde{r}_N) \delta(x-x_j),
\end{align}
where 
\begin{align}
  \label{eq:DGLAPmeasureDef}
  \left[\textrm{d}x_i \textrm{d}^2k_\perp^i \right]_N = & \frac{1}{(16\pi^3)^{N-1}}\left[\prod_{i=1}^N \textrm{d}x_i \textrm{d}^2k_\perp^i\right] \nonumber \\
  & \times \delta\left(1 - \sum_j x_j \right) \delta^{(2)}\left(\sum_jk_\perp^j\right)
\end{align}
is the measure associated with the $N\to N$ parton overlap encompassing momentum conservation \cite{Diehl:2000xz}. $j$ labels the active parton selected, $s_j$ is the flavour of the active parton, and $\hat{r}_N$ and $\tilde{r}_N$ are the momenta argument of the outgoing and incoming LFWFs respectively, boosted from the frame where the hadrons have respectively no transverse momentum to the symmetric frame used to define GPDs. They depend on the $(x_i,k_\perp^i)$, on $\xi$ and $\Delta_\perp$. Their exact expression can be found in ref. \cite{Diehl:2000xz}.
The important point here is to notice that the operator acts on a quark transiting between the initial and final state, thus selecting a diagonal configuration in terms of partons number as shown on fig. \ref{fig:OverlapPicture}.

A contrario in the ERBL region, the operator is contracted with a pair of quark-antiquark, extracting it from the initial state, as illustrated on fig. \ref{fig:OverlapPicture}. Consequently, in that case, the overlap is not diagonal and can be sketched as:
\begin{align}
  \label{eq:ERBLOverlap}
  \left. H^q(x,\xi)\right|_{x\le |\xi|} \propto & \sum_N \sqrt{1-\xi}^{2-N } \sqrt{1+\xi}^{-N }\nonumber \\
  & \times \sum_{j,j'} \frac{\delta_{\bar{s}_j' s_j}\delta_{s_j q}}{\sqrt{n_j n_{j'}}}\int \left[\textrm{d}x_i \textrm{d}^2k_\perp^i \right]_{N-1}^{N+1} \nonumber \\
  & \times\left(\Phi^{N-1} (\hat{r}_{N-1})\right)^* \Phi^{N+1}(\tilde{r}_{N+1}) \delta(x-x_j),
\end{align}
where again $\left[\textrm{d}x_i \textrm{d}^2k_\perp^i \right]_{N-1}^{N+1}$ is the appropriate measure associated with the $N+1 \to N-1$ transition and encompassing momentum conservation:
\begin{align}
  \label{eq:ERBLmeasureDef}
  \left[\textrm{d}x_i \textrm{d}^2k_\perp^i \right]_{N-1}^{N+1} & = \textrm{d}x_j \prod_{i\neq j,j'}^{N+1} \textrm{d}x_i \delta \left(1-\xi - \sum_{i\neq j,j'}^{N+1}x_i \right) \nonumber \\
                                                                & \quad \times \frac{\textrm{d}^2k_\perp^j}{(16\pi^3)^{N-1}}  \prod_{i\neq j,j'}^{N+1} \textrm{d}k_\perp^i \nonumber \\
                                                                & \quad \times \delta\left(\frac{\Delta_\perp}{2}-\sum_{i\neq j,j'}k_\perp^i \right).
\end{align}
In this area, the absence of parton number conservation between the initial and final state makes any attempt to compute GPDs from lightfront wave functions ambiguous, as one needs to truncate the Fock space at a given order $N$. This has led to models built in the DGLAP region only \cite{Mezrag:2016hnp,Rinaldi:2017roc,Shi:2020pqe}. We will see in sec. \ref{sec:CovExt} how to bypass this difficulty.

\subsection{The positivity property}

Before tackling the difficulty of the ERBL region presented above, we focus on the simplification brought by the specific case of the forward limit, \emph{i.e.} when $\Delta \to 0$ and thus the matrix element becomes diagonal in momentum space. In that case, there is no need to boost the incoming or outgoing hadron and we are left with:
\begin{align}
  \label{eq:PDFOverlap}
  q(x) \propto & \sum_N \sum_j \delta_{s_j,q}\int \left[\textrm{d}x_i \textrm{d}^2k_\perp^i \right]_N \nonumber \\
  & \times\left|\Phi^N (x_i,k_\perp^i)\right|^2 \delta(x-x_j).
\end{align}
The PDF is thus obtained by summing the modulus square of the LFWFs over all possible momenta and partons content. We recover here the probabilistic aspect of the PDF (number density) with respect to the LFWFs acting as regular quantum mechanical coefficients weighting Fock states in a Fock basis. Formally, we thus obtain the PDF as the norm of a vector belonging to a Hilbert space.

Such a formal structure is not restricted to the forward limit, but can be extended to the entire DGLAP region. Since in this region the overlap conserve the parton number, then one can write the GPDs formally as a scalar product:
\begin{align}
  \label{eq:GPDasScalProd}
  \left. H(x,\xi)\right|_{x\ge \xi} = \bra{\Phi_{\textrm{out}}}\left. \Phi_{\textrm{in}}\right\rangle.
\end{align}
Using the Cauchy-Schwartz inequality, we obtain:
\begin{align}
  \label{eq:CauchySchwartz}
    |\left. H(x,\xi)\right|_{x\ge \xi}| \le || \ket{\Phi_{\textrm{in}}}|| \times ||\ket{\Phi_{\textrm{out}}}||.
\end{align}
The forward limit then tells us how to relate the vector norms with the PDF. However, its momentum dependence needs to be boosted from the incoming and outgoing  hadron frames, in which the hadron has no transverse momentum. The procedure yields the so-called positivity condition \cite{Radyushkin:1998es,Pire:1998nw,Diehl:2000xz,Pobylitsa:2002gw,Pobylitsa:2001nt}:
\begin{align}
  \label{eq:Positivity}
  |\left. H^q(x,\xi)\right|_{x \ge \xi} | \le \sqrt{q\left(\frac{x-\xi}{1-\xi}\right) q\left(\frac{x+\xi}{1+\xi}\right)},
\end{align}
offering an upper and lower bounds on the GPD in the DGLAP region. Eq. \eqref{eq:Positivity} is given for a quark GPDs in a pion, but the same kind of inequality can also be derived for gluons, and within baryon with little modifications. Last but not least, we highlight that the LFWFs formalism, by explicitly relying on the Fock state structure, yields by construction models fulfilling the positivity in the DGLAP region. 

\subsection{Covariant extension}
\label{sec:CovExt}

So far, we have thus highlighted two modelling techniques intrinsically related to two major properties of GPDs: the DD approach allowing by construction to fulfil polynomiality, and the Lightfront Wave function formalism built so that positivity is automatically implemented. However, positivity is not manifest in the DD framework, and the difficulties related to the ERBL region make the fulfilment of polynomiality delicate at best in the LFWFs formalism. Therefore, physicists have historically favoured one property over the other, sticking to one types of modelling formalism.

This situation has changed in the last few years. First, let us mention that progresses in modelling techniques of GPDs in the DD space have very recently allowed the authors of ref. \cite{Dutrieux:2021wll} to systematically exclude DD models not fulfilling the positivity property. In a nutshell, the technique consists in exploring the DD phase space compatible with data, by generating replicas, and then disregarding the replicas not compatible with positivity. This technique has the advantage of strongly reducing the uncertainties on the extracted GPDs, but the drawback of necessitating a large amount of computing time. This is an issue, as in the end, many replicas as discarded.

Another technique has been proposed few years ago in ref. \cite{Chouika:2017dhe,Chouika:2017rzs} (see also ref. \cite{Hwang:2007tb} for a pioneering example on a different model), and exploits both the DD and the LFWFs frameworks. It consists in modelling the DGLAP region with LFWFs (or more generally in a way that conserve the scalar product structure) so that the positivity is fulfilled by construction. Then, from the DGLAP region only, we extract the underlying DDs by use of the inversion of the Radon transform. Finally, the ERBL region is computed from the DDs, so that polynomiality is automatically satisfied. Two questions arise from this: i) is the inverse Radon transform possible from the DGLAP region only and ii) if possible, is it unique?

We will proceed by giving an intuitive answer to such a question. Looking at fig. \ref{fig:DDsupport}, one notes that the integration lines corresponding to the DGLAP region cross the $\beta = 0$ outside the support domain. Any point of the domain, excepting the points on the $\beta=0$ line, belong to infinitely many DGLAP-type lines. Thus, the DGLAP region contains indeed all the necessary information to get the DDs back, except the D-term, which lives solely on the line $\beta=0$, and in general cannot be obtained by continuity (see eq. \eqref{eq:DDtoGPD})\footnote{In some specific DD schemes, $F$ and $G$ DDs are given by a unique distribution $h$. In that case, determining $h$ allows one to extract a $D$-term. However, this $D$-term is generally not unique, and scheme transformations may require the introduction of an additional regularising $D$-term. Technical details are given in ref. \cite{Chouika:2017dhe}.}. Moreover, we intuitively see that the small $\beta$ region will be harder to extract from the numerical point of view, as the DGLAP lines probing it becomes steeper and steeper.

\begin{figure}[t]
  \centering
  \includegraphics[width = 0.45\textwidth]{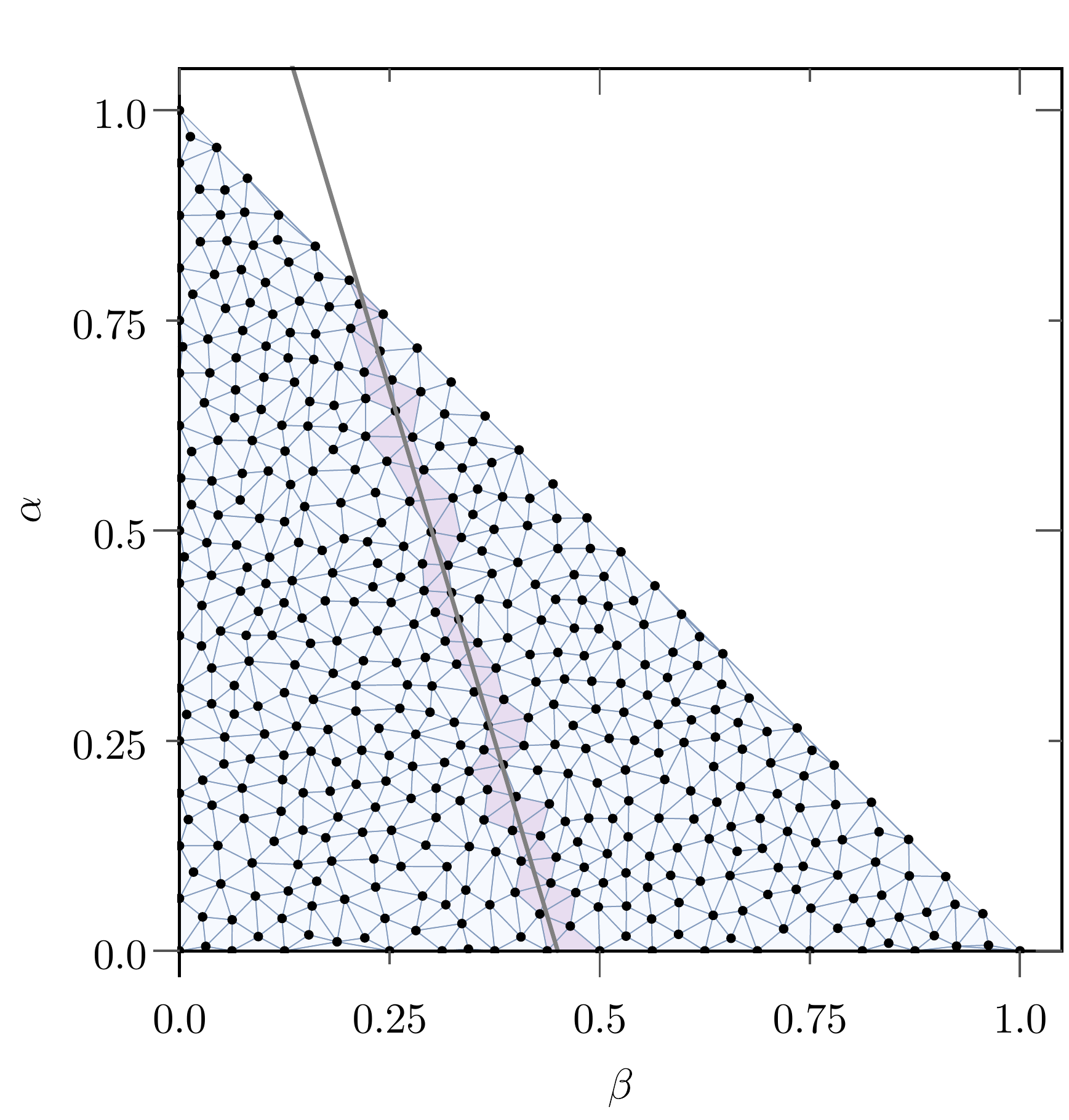}
  \caption{Discretisation of the DD support through a Delaunay mesh. A DGLAP line hit a number of cells. Being given a sufficient amount of lines, one can extract a Finite Element approximation of the Double Distribution. Figure from \cite{Chavez:2021llq}}
  \label{fig:DelaunayTriangulation}
\end{figure}

The uniqueness of the DD extracted from the DGLAP region only comes from the fact that DDs are compactly supported. Indeed, in this case, in can be shown from the Boman and Todd-Quinto theorem \cite{boman:1987rad}, that if a GPD is vanishing on the DGLAP region, (see ref. \cite{Chouika:2017dhe} for technical details), then the DD behind it is also zero. This leaves room only for a $D$-term, manifesting itself only in the ERBL region, and is consistent with the intuitive picture built before.  

Knowing that the solution is unique, one may try to derive it. However, since the inverse Radon problem is infamously ill-posed in the sense of Hadammard, one expects that numerical noise will need to be tamed. The choice which has been made to tackle this problem is to approximate the DD space through finite element methods \cite{Chouika:2017dhe,Chavez:2021llq}. Within this framework, the DD space is discretised into elements (see fig. \ref{fig:DelaunayTriangulation}) composed of vertices and cells. The DD is then approximated by polynomials of given degrees on each cell whose values is fixed at nodes, which may be shared with neighbouring cells, guaranteeing the continuity of the reconstructed function. The Radon inverse problem is then reduced to a matrix inverse problem such that:
\begin{align}
  \label{eq:discreteRadon}
   \begin{pmatrix}
    \\
    H_{i}\\
    \\
  \end{pmatrix}
  =
  \begin{pmatrix}
    & & \\
    & \mathcal{R}_{ij} & \\
    & & 
  \end{pmatrix}
  \begin{pmatrix}
    \\
    F_j\\
    \\
  \end{pmatrix},
\end{align}
where $H_i$ is the value of the GPD in the DGLAP region at a point $(x_i,\xi_i)$, $F_j$ are the value of the DD at the nodes, and $\mathcal{R}_{ij}$ is the discretised Radon transform matrix of the basis. The interested reader can find numerical details in ref. \cite{Chavez:2021llq}. We will only highlight that through this method, the system converges toward a solution, providing that the rank of the matrix $\mathcal{R}$ is maximal. 
Moreover, over-constraining the system and solving it through normal equations allow us to improve the conditioning of the matrix, and therefore to tame numerical noise. Systematic uncertainties can be assessed through multiple sampling of the DGLAP lines and the use of multiple discretisation grids.

Finally, let us mention that this new technique has been benchmark on algebraic models of the GPDs \cite{Chouika:2017rzs} and has been used to produce prediction of DVCS cross-sections for the future electron-ion colliders in the USA and in China \cite{Chavez:2021koz}. However, no attempt to extract GPDs from experimental data has been performed so far using this new technique. An alternative discussion using the Laplace transform has been developed in \cite{Muller:2017wms}, but never applied on concrete cases to the best of my knowledge.


%% file: section_evolution.tex
We have highlighted in the previous section the main properties that GPDs need to fulfil. But we have avoided until now a major topic of quantum field theory, namely renormalisation. In this section, we will present the renormalisation group properties of GPDs and discuss two ways to compute their scale dependence. This section is largely inspired by ref. \cite{Bertone:2022frx}, and we invite the interested reader to consult it for detailed computations and discussions about the evolution properties of GPDs. Since the $t$-dependence of GPD do not play any role in their evolution, we will drop it in our notations in this section.

\subsection{Renormalisation of local operators}

Looking back at eq. \eqref{eq:HqDef}, the non-local operator \linebreak
$\bar{\psi}^q(-z/2)\gamma^+ \psi^q(z/2)$ entering the definition of the matrix element presents typical quantum field theory short-distance singularities when $z\to 0$. This tells us that the operator needs to be renormalised, so that the short-distance singularities are properly regularised and traded for a scale dependence. We emphasise for readers unfamiliar with renormalisation of composite operators that it is not sufficient to renormalise the quark fields themselves through the canonical $\psi_R = Z_2^{-\frac{1}{2}} \psi $ renormalisation procedure, but that a genuine singularity from the operator itself needs to be tamed. Details are given in classical textbooks such as ref. \cite{Peskin:1995ev}.

Two classical ways can then be exploited to treat these singularities. The first one, called operator product expansion (OPE) consists formally into Taylor expanding the non-local operator into local ones labelled $O_n$:
\begin{align}
  \label{eq:OPE}
  \bar{\psi}^q(-z/2)\gamma^+ \psi^q(z/2) = \sum_n c_n(z) O_n(0).
\end{align}
The $c_n(z)$ coefficients contain short-distance physics and can thus be computed using perturbation theory. The local operators $O_n$ need to be renormalised, and one defines renormalisation constants $Z_{O_n}$ so that:
\begin{align}
  \label{eq:OPERenorm}
  O_n^R = Z_{O_n}O_n.
\end{align}
Thus, it is necessary in principle to provide a renormalisation condition for each local operator. In such a context, the  $\overline{\mbox{MS}}$ scheme provide a systematic provide, although inseparable from dimensional regularisation, to provide renormalisation conditions for the entire tower of operators. 

Introducing the renormalised local operators $O_n^R$ in eq. \eqref{eq:OPERenorm}, we omitted a typical feature of QCD: operators usually mix with each others, and this mixing includes quarks operators with gluons operators. However, it is possible to decompose quark operators between specific combination called singlet and non-singlet, respectively mixing and not mixing with the gluons operators.
These combinations were already introduced in sec. \ref{sec:GPDs} when we discussed the $x$-parity of quarks GPDs.
We thus remind the reader that the quark flavoured singlet GPD $H^+_q$ introduced in eq. \eqref{eq:SingletGPD} are odd in $x$, while the non-singlet combinations $H^-_q$ introduced in eq. \eqref{eq:NonSingletGPD} are even.
However, when looking at the coupling between quarks and gluons, one should not consider $H_q^+$ for a given flavour $q$, but rather the sum over all active flavours at the considered scale. This is illustrated in the next section.

One can even further simplify the mixing. Indeed, in the non-singlet case, perturbative renormalisation at leading-order is such that specific local operators, called conformal operators, do not mix with each other under renormalisation. These operators can be obtained by computing the moment of the non-singlet GPDs with the $C_n^{(3/2)}$ Gegenbauer polynomial:
\begin{align}
  \label{eq:Gegenbauer}
  O_n(\xi,t) = \xi^n \int_{-1}^1 \textrm{d}x\, C_n^{(3/2)}\left(\frac{x}{\xi}\right) H_q^-(x,\xi,t).
\end{align}
Because of the $x$-parity of $H^-$, only even $n$ are non-vanishing. These moments evolve multiplicatively at leading order, simplifying greatly the description of evolution. However, the difficulty lies in reconstructing the $x$-dependence from a finite number of conformal moments. Polynomials reconstructions are usually slow to converge, and thus resummation techniques based on Mellin-Barne transform have been introduced \cite{Mueller:2005ed}. Moreover, at higher order of perturbation theory, the conformal moments start mixing, necessitating to find a new eigen basis of the evolution operator\footnote{An alternative is to use the so-called conformal-subtraction scheme, removing the mixing of operators at NLO \cite{Kumericki:2007sa}.}. Despite these difficulties, an evolution code in conformal space exists \cite{Kumericki:2007sa} and has been use for phenomenological studies \cite{Mueller:2013caa,Kumericki:2016ehc}.

\subsection{Renormalisation of light-ray operators}

\begin{figure}[t]
  \centering
  \includegraphics[width=0.5\textwidth]{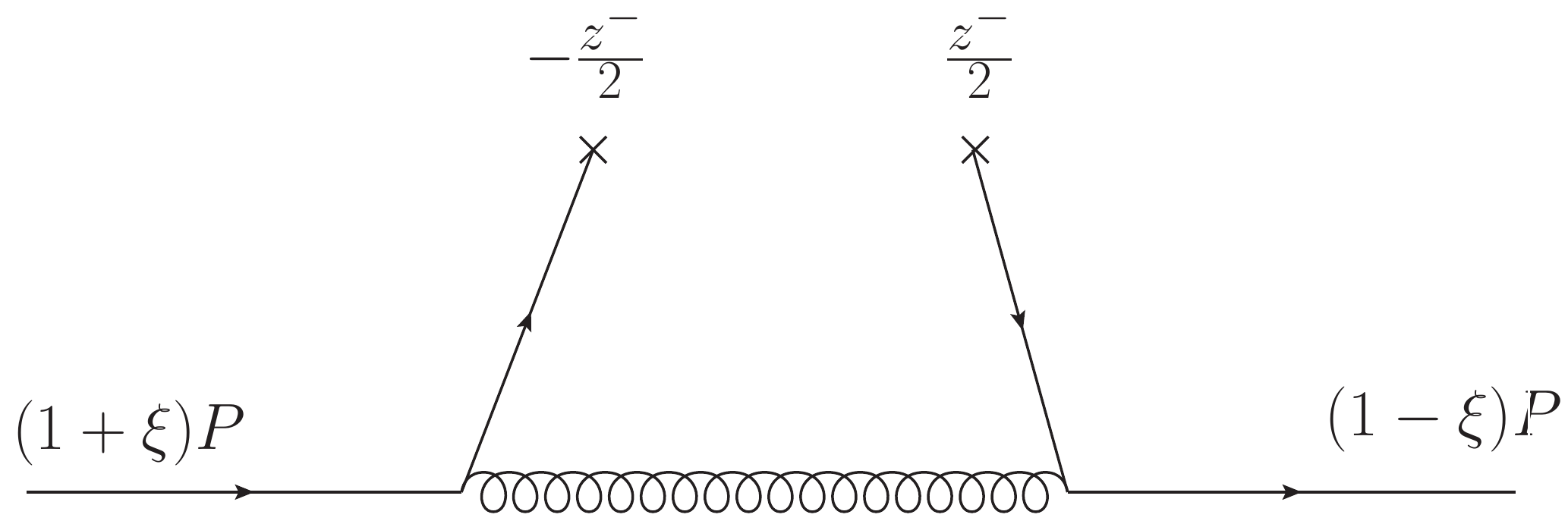}\\
  \includegraphics[width=0.5\textwidth]{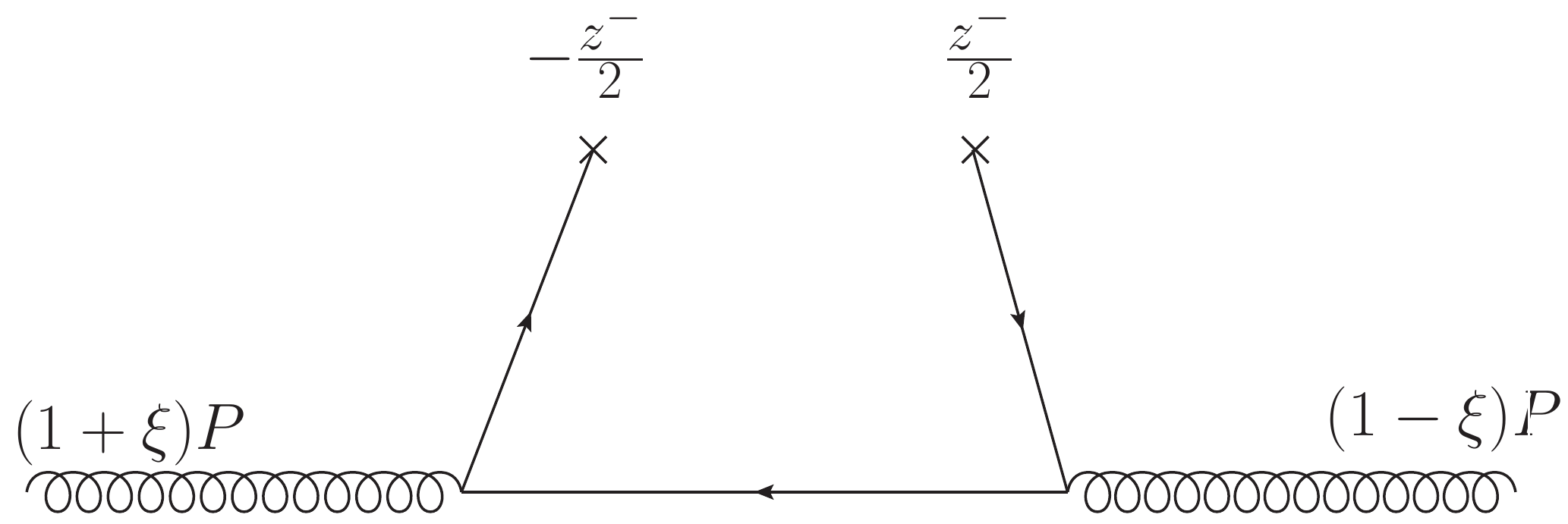}
  \caption{Two leading order diagrams contributing to the computation of $Z$. Top: quarks-in-quarks GPD; Bottom: quarks-in-gluons GPD illustrating the mixing under evolution.}
  \label{fig:GPDEvolution}
\end{figure}

Alternatively, one can directly renormalise the $x$-dependent GPDs by defining an $x$-dependent renormalisation function $Z$, allowing us to bypass the expansion on conformal operators, and the necessary reconstruction following it. We sketch the procedure which is greatly detailed in ref. \cite{Bertone:2022frx}. One can formally relate the scale-dependent GPD $H(\mu)$ and the regularised GPD $\hat{H}(\epsilon)$ through:
\begin{align}
  \label{eq:RenormalisedGPD}
  H^i(x,\xi,\mu) & = \int_{-1}^1 \frac{\textrm{d}y}{|y|} Z^{ij}\left(\frac{x}{y},\frac{\xi}{x},\alpha_s(\mu),\epsilon \right) \hat{H}^j(y,\xi,\epsilon),
\end{align}
where $(i,j)$ label quarks or gluons contributions. In the $\overline{\mbox{MS}}$ scheme, the computation of $Z$ only requires to study the UV behaviour of the operator and the associated poles in dimensional regularisation. This pole structure is independent of the type of the nature of the incoming and outgoing states considered, allowing us to compute $Z$ exploiting partons-in-partons GPDs (see fig. \ref{fig:GPDEvolution}).

Eq. \eqref{eq:RenormalisedGPD} tells how to trade the regulator dependence of the GPD for the scale dependence. However, for phenomenological purposes, we are more interested in knowing how the GPD is modified by changing the scale at which it is computed. To do that, we would like to know how $H(\mu + \textrm{d}\mu)$ is modified with respect to $H(\mu)$. We can exploit the fact that $\hat{H}$ is independent of $\mu$ and write down the equivalent Callan-Symanzik equation for GPDs, called the evolution equation. In the non-singlet case, it yields:
\begin{align}
  \label{eq:EvolutionNS}
  \frac{\textrm{d}H_q^-(x,\xi,\mu)}{\textrm{d}\ln (\mu)} = \frac{\alpha_s(\mu)}{4\pi} \int_x^\infty \frac{\textrm{d}y}{y} & \mathcal{P}^{0-}_{q\leftarrow q} \left(y, \frac{\xi}{x}\right) \nonumber \\
  & \times  H_q^-\left(\frac{x}{y},\xi,\mu\right),
\end{align}
where $\mathcal{P}^0_{q\leftarrow q}$ is the leading order splitting function corresponding to a ``quark-in-quark'' GPD. Eq. \eqref{eq:EvolutionNS} is an order 1 differential equation with respect to $\alpha_s$, meaning that a leading order computation does not yield a power correction in the strong coupling, but rather an exponential one. However, we also note that contrary to conformal space evolution, it is necessary to numerically solve a differential equation to obtain the evolved GPDs. Depending on the final aim of a given computation, it can be more interesting to solve this differential equation than to inverse Mellin-Barne transform and reconstruct the $x$-dependence from the evolved conformal moments.

In the singlet sector, the situation is similar, albeit a bit more complex because of the mixing between quarks and gluons operators. One can write the evolution equation as:
\begin{align}
  \label{eq:EvolutionS}
    \frac{\textrm{d}H^+(x,\xi,\mu)}{\textrm{d}\ln (\mu)}
  = \frac{\alpha_s(\mu)}{4\pi} \int_x^\infty \frac{\textrm{d}y}{y} \mathcal{P}^0_{+} \left(y, \frac{\xi}{x}\right)\nonumber \\
  \times H^+\left(\frac{x}{y},\xi,\mu\right),
\end{align}
where:
\begin{align}
  \label{eq:HplusDef}
  H^+(x,\xi,\mu^2) =
  \begin{pmatrix}
    \sum_q H^+_q(x,\xi,\mu^2) \\
    H^g(x,\xi,\mu^2)
  \end{pmatrix}
  ,
\end{align}
and
\begin{align}
  \label{eq:PplusDef}
  \mathcal{P}^{0+} =
  \begin{pmatrix}
    \mathcal{P}^{0,+}_{q\leftarrow q} & \quad  2 n_f \mathcal{P}^{0,+}_{q\leftarrow g} \\
    \mathcal{P}^{0,+}_{g \leftarrow q} & \mathcal{P}^{0,+}_{g\leftarrow g}
  \end{pmatrix},
\end{align}
where $j\leftarrow i$ label the transition from partons type $i$ to parton type $j$ (see fig. \ref{fig:GPDEvolution} for an example of $q \leftarrow q$ and $q\leftarrow g$ cases). The expression of the LO splitting function within the present notation convention can be found in ref. \cite{Bertone:2022frx}. We highlight that LO splitting function have been obtained in the 1990s \cite{Mueller:1998fv,Ji:1996nm,Radyushkin:1997ki,Blumlein:1997pi}, next-to-leading order correction to evolution kernels followed quickly \cite{Belitsky:1998vj,Belitsky:1999fu,Belitsky:1999hf} and were recently confirmed \cite{Braun:2019qtp}. Moreover, computation at three loops in the non-singlet sector have been recently performed \cite{Braun:2017cih}. However, few evolution codes in $x$-space are available. One can mention the so-called Vinnikov code \cite{Vinnikov:2006xw} which is not maintained anymore, and the recently implemented GPD sector of Apfel++ \cite{Bertone:2013vaa,Bertone:2016lga,Bertone:2017gds,Bertone:2022frx}. These available implementations of evolution equations are limited to the LO splitting functions. Higher-orders evolution in $x$-space is nowadays highly desirable, notably for impact studies of the future electron-ion colliders.

\subsection{Properties of the GPDs evolution kernel}

The GPD evolution kernel and its associated splitting functions $\mathcal{P}^{0-}$ and $\mathcal{P}^{0+}$ presents interesting contributions that we highlight in the following. First, let us mention that the splitting functions can be decomposed into two functions $\mathcal{P}_1$ and $\mathcal{P}_2$ contributing into two different, but overlapping kinematic domains:
\begin{align}
  \label{eq:EvolutionDomaine}
  \mathcal{P}^{0 \pm}\left(y,\frac{\xi}{x}\right) = & \theta(1-y) \mathcal{P}_1^{0 \pm}\left(y,\frac{\xi}{x}\right) \nonumber \\
  & + \theta\left(\frac{\xi}{x}-1 \right) \mathcal{P}_2^{0\pm}\left(y,\frac{\xi}{x}\right).
\end{align}
This decomposition allows us to make a first remark: in the forward limit of GPDs, only $\mathcal{P}_1^{0 \pm}$ contributes and thus we expect that:
\begin{align}
  \label{eq:EvoDGLAPLimit}
  \lim_{\xi\to 0} \mathcal{P}_1^{0 \pm}\left(y,\frac{\xi}{x}\right) = P^0_{\textrm{DGLAP}}(y),
\end{align}
which is indeed verified. The $\mathcal{P}_2^{0\pm}$ part of splitting function is active only in the inner kinematic region $-\xi \le x \le \xi$, so that the kernel describes the evolution of a pair of quark-antiquark, similarly to meson distribution amplitudes. And in fact, when $\xi \to 1$ the GPDs evolution kernel reduces itself to the ERBL one. This justifies a posteriori the names DGLAP and ERBL given to the outer and inner kinematic regions.

In the ERBL region, we highlight that singularities appear in $\mathcal{P}_1^{0\pm}$ and $\mathcal{P}_2^{0\pm}$, when $y \to \frac{x}{\xi}$ but nicely cancel each other. Moreover, close to the $x = \xi$ line, one can prove that:
\begin{align}
  \label{eq:EvoContinuity}
  \mathcal{P}_2^{0\pm}(y,\frac{\xi}{x}) \propto 1 - \frac{\xi}{x},
\end{align}
guaranteeing the continuity of the splitting function at the crossover line. However, the splitting function is not differentiable at the crossover line, generating cusps when evolving smooth functions (see ref. \cite{Bertone:2022frx} for examples). This contrasts with the claim sometimes found in the literature that evolution makes GPDs smoother.

The ERBL and DGLAP limits can also be recovered in the conformal space formalism. Indeed, looking at the conformal moment definition in eq. \eqref{eq:Gegenbauer}, one realises in the limit $\xi \to 0$, the conformal moments of the GPD yield the Mellin moment of PDF. The latter are known not to mix among each other at all order of perturbation theory. In the limit $\xi \to 1$, the Gegenbauer polynomials form a basis on the entire $x$-space, diagonalising the GPD evolution kernel, consistently with its ERBL limit.

Last but not least, let us mention the physical constraints on the moments, coming from the local matrix elements of conserved currents. The scale independence of the electromagnetic form factor implies that the integral of $\mathcal{P}^{0,-}(y,\frac{\xi}{y x})$ over $y$ vanishes. The conservation of the total momentum implies that the integrals of $ y \mathcal{P}^{0,+}_{q \leftarrow q}(y,\frac{\xi}{y x})$ and of $ y \mathcal{P}^{0,+}_{g \leftarrow q}(y,\frac{\xi}{y x})$ compensate each other, and the same thing happens for the lower line of the matrix   $ \mathcal{P}^{0,+}$ .


%% file: section_exclusive.tex
Until now, we have mainly highlighted the theoretical properties of GPDs. In this section, we will focus on their experimental access, reviewing the main processes explored at past, current and future facilities. Since there is no pion target, this section will be mostly dedicated to the nucleon, with the brief exception of the discussion on the Sullivan process.

\subsection{DVCS}

\begin{figure}[t]
  \centering
  \includegraphics[width=0.45\textwidth]{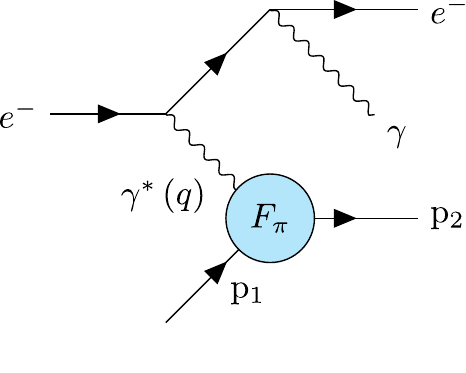}
  \caption{Bethe-Heitler process interfering with DVCS. The real photon can be emitted by the incoming or outgoing electron.}
  \label{fig:BHprocess}
\end{figure}

The most studied process allowing us to experimentally probe GPDs is certainly DVCS, mentioned in the introduction (see fig. \ref{fig:DVCS}). However, in the introduction, we have omitted that DVCS is not the only process such that $ep \to ep\gamma$. It interferes with the so-called Bethe-Heitler (BH) process, illustrated on fig. \ref{fig:BHprocess}, and which involves the nucleon form factor rather than the GPDs. Consequently, the cross-section of the $ep\to ep\gamma$ processes is the sum of the DVCS and BH amplitudes, interfering among each other. This yields a differential cross-section depending on three different terms (see ref. \cite{Kumericki:2016ehc} and references therein for details)
\begin{align}
  \label{eq:DVCSCrossSection}
  \frac{\mathrm{d}^5\sigma}{\textrm{d}x_B\textrm{d}Q^2\textrm{d}|t|\textrm{d}\phi \textrm{d}\phi_S} \propto & |\mathcal{T}_{\textrm{BH}}|^2 + |\mathcal{T}_{\textrm{DVCS}}|^2 + \mathcal{I}, 
\end{align}
where $\phi$ is the angle between the leptonic and hadronic planes and $\phi_S$ the angle between the leptonic plane and the transverse polarisation of the target. $\mathcal{T}$ designates the amplitude of pure DVCS or BH contributions, while $I$ stands for the interference terms. Usually\footnote{Alternative description relying on amplitudes are advocated in the literature \cite{Diehl:2003ny,Kriesten:2020wcx}.}, the amplitudes are expanded in harmonics of $\phi$ \cite{Belitsky:2001ns,Belitsky:2008bz,Belitsky:2010jw,Belitsky:2012ch} whose coefficients are themselves parameterised in terms of Compton form factors (CFFs). Any attempt to extract GPDs from experimental data starts with the extraction of these CFFs. It is already at the level of CFFs that some theoretical approximations are performed, such as the twist expansion.

The extraction of CFFs has been one of the major activities in the study of GPD-related processes in the past decade, both on the theoretical and phenomenological side. Very recently, it has taken a new turn with the combination of new experimental data released on the neutron \cite{Benali:2020vma}, but also with the use of machine learning techniques and in particular artificial neural networks (ANN) \cite{Cuic:2020iwt,Moutarde:2019tqa}. Theoretical progresses in the last decade have also brought a refined parameterisation of the amplitude, considering higher-twist components, such that the kinematic corrections in $t/Q^2$ published in refs. \cite{Braun:2012hq,Braun:2012bg} and exploited in ref. \cite{Defurne:2015kxq}. From these pioneering studies (and as suspected before), one expects that the low-$Q^2$ data coming from various facilities such as Jefferson Laboratory contain significant higher-twist contributions.

\subsection{The deconvolution problem and Shadow GPDs}

Once extracted, the CFF $\mathcal{H}$ still needs to be expressed in terms of GPDs. They are given as a convolution of GPDs with a hard scattering kernel $T$ which describes the interaction of the active parton (or a pair of partons in the ERBL region) with the highly virtual and the real photons. Formally, it yields:
\begin{align}
  \label{eq:CFFandGPD}
  \mathcal{H}^{q}(\xi,t,Q^2) =  \int_{-1}^1 \frac{\textrm{d}x}{2\xi} &T^{q}\left(\frac{x}{\xi},\frac{Q^2}{\mu^2},\alpha_s(\mu^2) \right) \nonumber \\
  & \times H^{+}_q(x,\xi,t,\mu^2),
\end{align}
where $T^q$ is the quark hard scattering kernel that can be expanded in perturbation theory thanks to its dependence in the hard scale $Q^2$. And, in fact, one-loop corrections have been derived since the end of the 1990s \cite{Mankiewicz:1997bk,Ji:1998xh,Belitsky:1999sg,Freund:2001hd,Freund:2001rk,Pire:2011st} while two loops corrections have been recently made available \cite{Braun:2020yib}. However, NLO studies of DVCS have been rather limited, despite the pioneering approaches in refs. \cite{Kumericki:2007sa,Moutarde:2013qs} and the recent attempt in ref. \cite{Chavez:2021llq,Chavez:2021koz}.

Eq. \eqref{eq:CFFandGPD} triggers immediately the following question: is it possible to recover the GPD from the CFF despite the convolution ? This question is known as the DVCS deconvolution problem. It was addressed at the end of the 1990s \cite{Freund:1999xf} and commonly believed thereafter that evolution would be the keystone of the deconvolution procedure. Indeed, since at LO the imaginary part of $T$ selects only the configurations such that $x = \pm \xi$, it is obvious that considering only a leading order expansion of $T^q$, would not allow extracting GPDs from DVCS.
However, the question remained open at higher-orders, as it was unclear whether the combination of probing the entire GPD support together with the theoretical constraint on GPDs would be sufficient to allow an unambiguous extraction, at least in principle.

The answer was brought by the recent study of ref. \cite{Bertone:2021yyz}. Therein, the concept of shadow GPD $H_0^{(n)}$ of order $n$ was introduced and defined so that at a given scale $\mu_0$:
\begin{align}
  \label{eq:ShadowDefDIS}
  0 & = H_0^{(n)}(x,0,0,\mu_0), \\
  \label{eq:ShadowDefDVCS}
  0 & = \mathcal{H}^{(j)}(\xi,t,\mu_0^2)  =  \int_{-1}^1 \frac{\textrm{d}x}{2\xi} T^{q,(j)}\left(\frac{x}{\xi},1,\alpha_s(\mu_0^2) \right)\nonumber \\
  & \quad \hspace{3cm} \times H^{(j)}_0(x,\xi,t,\mu_0^2), \textrm{ for } j\le n
\end{align}
where $ T^{q,(j)}$ is the DVCS perturbative kernel expanded at order $j$ in $\alpha_s$.
Consequently, if it exists, $H_0^{(n)}$ is invisible both in DIS (eq. \eqref{eq:ShadowDefDIS}) and in DVCS described up to order $n$ of perturbation theory (eq. \eqref{eq:ShadowDefDVCS}). Ref. \cite{Bertone:2021yyz} presents an algorithm allowing to systematically compute and determine such a GPD model considering all the properties of GPDs but positivity, by working with polynomials DDs of maximal degree $N$. One can define a threshold $N_{\textrm{Thres}}^{(n)}$ which depends on the perturbative order considered and such that when exploring polynomials DD space whose degree is larger than $N^{(n)}_{\textrm{Thres}}$, solutions of eq. \eqref{eq:ShadowDefDIS} and \eqref{eq:ShadowDefDVCS} can be found simultaneously. As expected, $N^{(n)}_{\textrm{Thres}}$ grows with higher orders of pQCD ($N^{(0)}_{\textrm{Thres}}= 9 $ while $N^{(1)}_{\textrm{Thres}}= 25$), meaning that the higher the perturbative order, the wider the functional space explored without encountering the shadow GPD problem.  

\begin{figure}[t]
  \centering
  \includegraphics[width=0.45\textwidth]{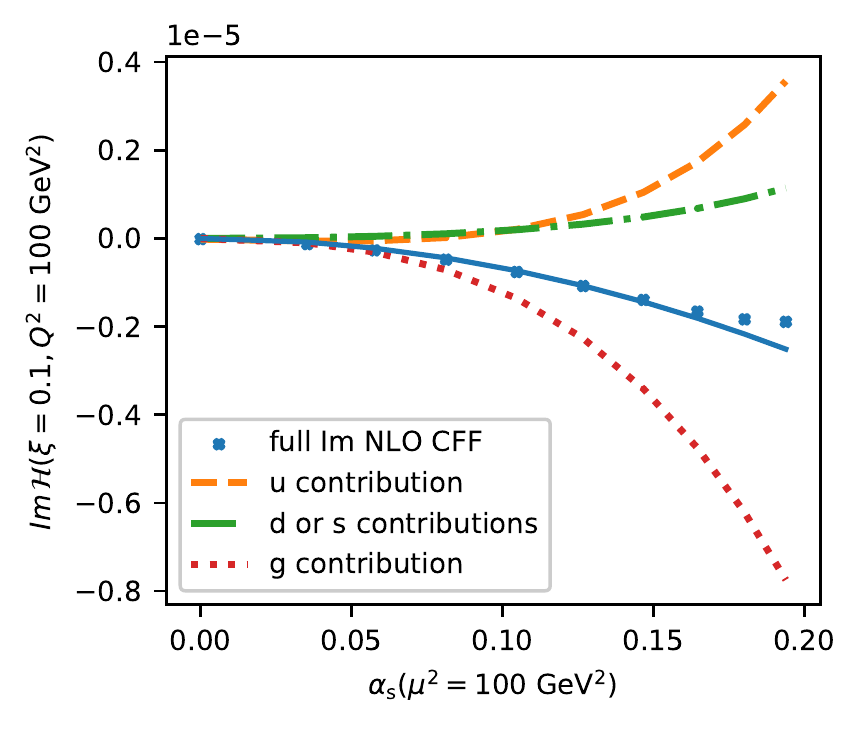}
  \caption{Behaviour of the imaginary part of $\mathcal{H}^{(1)}$ as a function of $\alpha_s$ after evolving a typical shadow GPD $H^{(1)}_0$ from 1 to 100 GeV$^2$. The blue dots are obtained for a set of value of $\alpha_s$ and the blue curve is a quadratic fit using seven points. Figure taken from ref. \cite{Bertone:2021yyz}.}
  \label{fig:EvolShadow}
\end{figure}

Regarding the impact of evolution, one expects that it remains limited. Indeed, the functions $H_0^{(1)}$ constructed in ref. \cite{Bertone:2021yyz} are such that when one writes
\begin{align}
  \label{eq:KernelExpansion}
  T^{(1)} = C_0 + \alpha_s(\mu) C_1 + \alpha_s(\mu^2) \ln\left(\frac{Q^2}{\mu^2} \right)C_{\textrm{coll}}
\end{align}
the convolution of $H_0^{(1)}$ with all $C_i$ coefficients vanishes.
Moreover, the evolution operator $\Gamma(\mu,\mu_0)$ defined such that:
\begin{align}
  \label{eq:evolutionoperator}
  H(\mu^2) = \Gamma(\mu^2,\mu_0^2) \otimes H(\mu_0^2),
\end{align}
where $\otimes$ designates the convolution over $x$,
admits the following $\alpha_s$ expansion:
\begin{align}
  \label{eq:evoexpansion}
  \Gamma(\mu^2,\mu_0^2) = 1 + \alpha_s(\mu) K^{(0)} \ln\left(\frac{\mu}{\mu_0} \right) + O(\alpha_s^2).
\end{align}
Therefore, one directly gets:
\begin{align}
  \label{eq:ScaleDependence}
   & \frac{\textrm{d}\mathcal{H}}{\textrm{d}\ln(\mu^2)}   = \frac{\textrm{d}T^{(1)}}{\textrm{d}\ln(\mu^2)} \otimes H + T^{(1)}\otimes \frac{\textrm{d}H}{\textrm{d}\ln (\mu^2)} \nonumber \\
   & = \alpha_s(\mu^2) C_{\textrm{coll}} \otimes H(\mu_0^2) + \alpha_s(\mu^2) C_0 \otimes K^{(0)} \otimes H(\mu_0^2)\nonumber \\
  & \quad + O(\alpha_s^2),
\end{align}
and from the scale independence of the CFF, we see that:
\begin{align}
  \label{eq:EqualityKernels}
  C_{\textrm{coll}} = - C_0 \otimes K^{(0)}.
\end{align}
This result has a major impact on the evolution impact on shadow GPDs. Indeed, computing the CFFs after evolving the shadow GPDs yields:
\begin{align}
  \label{eq:ShadowEvolution}
  \mathcal{H}^q(\xi,Q^2) & = T^{(1)} \otimes \left(1 + \alpha_sK^{(0)}\ln \left(\frac{\mu}{\mu_0} \right) \right) \otimes H_0^{(1)}(\xi,\mu_0^2) \nonumber \\
                         & = T^{(1)} \otimes H_0^{(1)}(\xi,\mu_0^2) + \alpha_s(\mu) \ln \left(\frac{\mu}{\mu_0} \right)\nonumber \\
                         &  \quad  \times C_0 \otimes K^{(0)} \otimes  H_0^{(1)}(\xi,\mu_0^2) + O(\alpha_s^2) \nonumber \\
                         & = O(\alpha_s^2).
\end{align}
Indeed, the first term cancels by definition of the shadow GPDs at scale $\mu_0$, while the second term cancels due to eq. \eqref{eq:EqualityKernels}. Thus, NLO shadow GPDs contribute to the CFF at order $\alpha_s^2$ up to large logarithms. 
Such a behaviour was numerically verified after evolving an example of shadow GPD $H^{(1)}_0$ from $\mu_0^2 = 1~\textrm{GeV}^2$ to $\mu^2 = 100~\textrm{GeV}^2$ as illustrated on fig. \ref{fig:EvolShadow}. We also note on the figure the order of magnitude of the generated CFF: $10^{-5}$ when the GPD itself is of order 1. Consequently, contributions from shadow GPDs to the CFF are not only formally speaking higher order in pQCD, but also extremely small, precluding us to rely solely on evolution to unambiguously extract GPDs from DVCS.

\subsection{Dispersion relations: Accessing the D-term}

Since the extraction of GPDs is strongly complicated by the existence of shadow GPDs, one can wonder whether it would be possible to access information on the nucleon structure from the CFF directly. The answer is yes, as it has been shown that dispersion relations exist between the real and imaginary parts of the DVCS CFFs. At leading order, the most popular dispersion relation is given by \cite{Anikin:2007yh}:
\begin{align}
  \label{eq:DispRel}
  \Re(\mathcal{H}(\xi,t,Q^2))  & = \int_{-1}^1 \frac{\textrm{d}x}{\pi}\Im (\mathcal{H}(x,t,Q^2))\nonumber \\
  & \times \left[\frac{1}{\xi-x}-\frac{1}{\xi+x} \right]  + S(t,Q^2)
\end{align}
where $S$ is the subtraction constant (with respect to $\xi$).
It allows one to derive the LO relation between $S$ and the $D$-term that has already been encountered in sec. \ref{sec:Polynomiality}:
\begin{align}
  \label{eq:SubtractionConst}
  S(t,Q^2) = 2 \int_{-1}^1 \textrm{d}\alpha \frac{D(\alpha,t,Q^2)}{1-\alpha}
\end{align}
The singularities in eq. \eqref{eq:DispRel} are regularised through the Cauchy principal value prescription. As we mentioned before, the D-term is related to the form factor $C$ of the EMT, which itself contains the information on the total pressure and the pressure anisotropy within the nucleon (see sec. \ref{sec:GPDs}). Eq. \eqref{eq:DispRel} has thus created the hope that one could extract information on the EMT through dispersion relations without handling the DVCS deconvolution problem.

To do so, it is first necessary to extract from experimental data the real and imaginary parts of the CFFs. Then, one needs to extrapolate the imaginary part to small values of $\xi$, so the integral in eq. \eqref{eq:DispRel} can be performed. This task is already challenging by itself. Extraction relying on ANN, thus reducing the parametrisation bias, are available in the literature \cite{Kumericki:2019ddg,Dutrieux:2021nlz}.
They yield subtraction constants which are compatible with 0 within error bands. Therefore, one can conclude that current data are not precise enough to exploit dispersion relations.

However, this may change with future facilities, notably the electron-ion collider which should provide a large amount of precise data on a wide kinematic range, especially in $Q^2$. The size of the $Q^2$ range is critical, since here evolution has a strong impact. Indeed, since the D-term only lives in the ERBL, not only its conformal moments do not mix under evolution (see sec. \ref{sec:Evolution}), but the Gegenbauer polynomials are a basis of such a subspace. The extraction of the conformal moments of the D-terms is thus in principle possible, thanks to the $Q^2$ dependence of the subtraction constant. More precisely, one can expand the D-term on the 3/2-Gegenbauer polynomial basis for quarks contributions:
\begin{align}
  \label{eq:DtermExp}
  D^q(z,t,Q^2) = (1-z^2) \sum_{n \textrm{odd}}^\infty d_n^q(t,Q^2) C_n^{3/2}(z),
\end{align}
and a 5/2-Gegenbauer polynomials basis for gluons:
\begin{align}
  \label{eq:DtermExpgluons}
  D^g(z,t,Q^2) = \frac{3}{2}(1-z^2)^2 \sum_{n \textrm{odd}}^\infty d_n^g(t,Q^2) C_{n-1}^{5/2}(z).
\end{align}
One can then realise that at LO, the subtraction constant is given by
\begin{align}
  \label{eq:Subtractiondcoef}
  S^q(t,Q^2) = 4 \sum_{n \textrm{odd}}^\infty d_n^q(t,Q^2).
\end{align}
A similar result can be obtained in the gluon sector. Eq. \eqref{eq:Subtractiondcoef} highlights very well the key role of evolution. In general, it would be impossible to disentangle $d_1$ from the rest of the coefficients of the series. But because the $Q^2$-dependence is controlled by the evolution equations and is different for every coefficient, we can in theory extract these coefficients provided that a large enough range in $Q^2$ is accessible. The interested reader will find the LO evolution of these coefficients in ref. \cite{Dutrieux:2021nlz}.
We recall here that since the D-term belongs to the singlet sector, quarks and gluons contributions mix under evolution. Moreover, a complete flavour separation between $u$, $d$ and $s$ quarks remains impossible. 

Finally, let us mention that we have provided here a LO analyses of dispersion relations, but that one can perform the same kind of analyses at higher order in perturbation theory. In such a case, eq. \eqref{eq:DispRel}, \eqref{eq:SubtractionConst} and \eqref{eq:Subtractiondcoef} are modified \cite{Diehl:2007jb}. No phenomenological analysis of NLO dispersion relations for DVCS has been published so far, but one has been performed and is currently in preparation for publication \cite{Meisgny}.

\subsection{Meson targets: The Sullivan process}

As we mentioned in the introduction of this section, most of the phenomenological studies are dedicated to the nucleon. However, with the future electron-ion collider, it may be possible to extend the DVCS studies from the nucleon to the pion through the so-called Sullivan process \cite{Sullivan:1971kd}. It consists in scattering a probe on a virtual pion within the nucleon.
Such an interpretation is expected to be valid if
\begin{itemize}
\item pion quantum number can be exchanged in the $t$-channel of the considered process,
\item $|t|$ is small.
\end{itemize}
In that case, one can expect that when $t\to m_\pi^2$, the cross-section of the considered process be schematically given as:
\begin{align}
  \label{eq:Sullivan}
  \frac{\textrm{d}\sigma^{\textrm{tot}}}{\textrm{d}t} & \propto \frac{1}{m_\pi^2 -t} \frac{\textrm{d}\sigma^{\textrm{pion pole}}}{\textrm{d}t} + \frac{\textrm{d}\sigma^{\textrm{reg}}}{\textrm{d}t}
\end{align}
where $\frac{\textrm{d}\sigma^{\textrm{reg}}}{\textrm{d}t}$ is regular for $t\to m_\pi^2$. The Sullivan idea has been used in several studies for determining the pion EFF \cite{Bebek:1977pe,JeffersonLabFpi:2000nlc,Huber:2008id} through notably the exclusive process $ep \to e \pi^+ n$, but also for extracting the PDFs of the pion through tagged DIS \cite{Barry:2018ort,Barry:2021osv} $ep\to enX$.

\begin{figure}[t]
  \centering
  \includegraphics[width=0.45\textwidth]{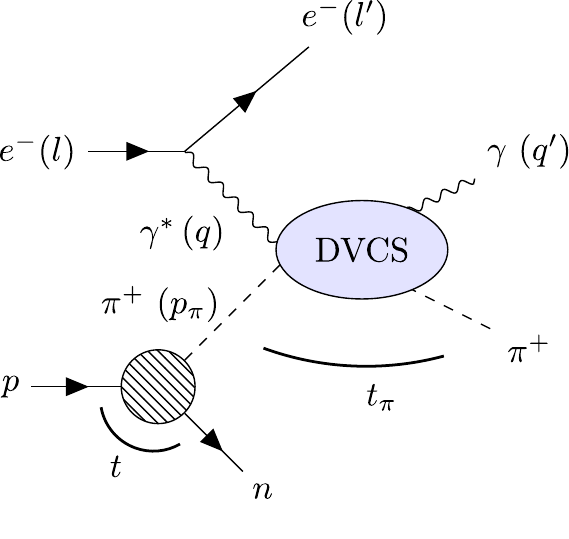}
  \caption{Sullivan-DVCS process. The deeply virtual photon interacts with a slightly virtual pion of the nucleon pion cloud. }
  \label{fig:SullivanDVCS}
\end{figure}

The case of tagged DVCS has been much less explored until now. Illustrated on fig. \ref{fig:SullivanDVCS}, it consists in interpreting the exclusive process $ep \to en\pi^+\gamma$ at low values of $|t|$ as a DVCS process on a virtual pion. From fig. \ref{fig:SullivanDVCS}, one notes that such interpretation is model-dependent, as the proton-neutron-pion vertex is needed to describe the process. Nevertheless, such a measurement remains today the only one that may bring experimental knowledge on pion GPDs.

Similarly to the nucleon case, the Sullivan-DVCS process interferes with the Sullivan-BH one. But other processes, related for instance to nucleon resonances, may also yield a similar final state. It is therefore necessary to carefully select the phase space so that resonances contributions can be neglected. We already highlighted the small-$|t|$ cut, but one also needs to cut on the invariant mass of the $n\pi$ system. Such additional cuts make the phase space allowed for Sullivan-DVCS measurement very challenging at current facilities (especially JLab 12) \cite{Amrath:2008vx}, but may allow us to extract pion GPDs at future facilities, such as the electron-ion colliders \cite{Chavez:2021llq,Chavez:2021koz}. Let us also mention that if feasible, the Sullivan-DVCS process will allow us to probe the structure of a \emph{virtual} pion, and thus, virtuality effects may need to be considered to recover the structure of a real pion \cite{Qin:2017lcd,Perry:2018kok}.

\subsection{Timelike Compton Scattering}

\begin{figure}[t]
  \centering
  \includegraphics[width=0.4\textwidth]{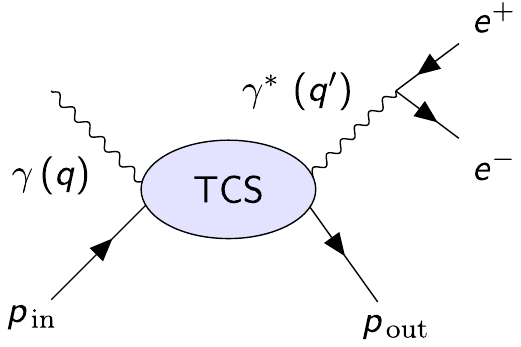}
  \caption{Timelike Compton Scattering. A real photon hits the nucleon which then emits a deeply virtual photon. This photon then decay into a lepton pair which can be detected at current and future facilities.}
\label{fig:TCS}
\end{figure}

Until now, we have only mentioned DVCS, but other exclusive processes are also related to GPDs. One of them is Timelike Compton Scattering (TCS) \cite{Berger:2001xd} illustrated on fig. \ref{fig:TCS} and in which the nucleon is hit by a real photon and emits a deeply virtual one, guaranteeing the existence of a hard scale in the process. This virtual photon then decay into a electron-positron pair that can be detected at existing facilities. Similarly to DVCS, TCS interferes with the BH process.

The first experimental signal related to TCS has been recently obtained at Jefferson Lab \cite{CLAS:2021lky}. Its similarity with DVCS make it the simplest exclusive process to exploit to check the universality of GPDs.
Indeed, its amplitude can be deduced from the one of DVCS, and thus theoretical developments performed for DVCS can be extended to TCS. However, because of this similarity, it shares the \emph{same} type of shadow GPDs than DVCS and thus does not provide a significant advantage regarding the deconvolution problem.

Already at the level of amplitudes, one can simultaneously exploit DVCS and TCS to check universality of GPDs. Indeed, as presented in ref. \cite{Grocholski:2019pqj}, analytic properties of DVCS and TCS amplitudes allows one to relate them at leading twist. An independent extraction of both DVCS and TCS amplitudes would thus allow us to check the universality through these specific relations (providing that leading twist dominance is guaranteed on the domain experimentally probed).

Finally, let us mention that TCS opens the possibility to extract GPDs at hadron colliders, in the specific case of ultra-peripheral collisions \cite{Lansberg:2015kha}. In these types of collisions, the impact parameter of the two hadrons is large enough so that they only interact through electromagnetism. One of the two becomes a source of quasi-real photons for the other one, allowing us in principle to measure TCS cross-sections. Feasibility studies are ongoing at the Large Hadron Collider (LHC). 

\subsection{Deep Virtual Meson Production}

\begin{figure}[b]
  \centering
  \includegraphics[width = 0.45\textwidth]{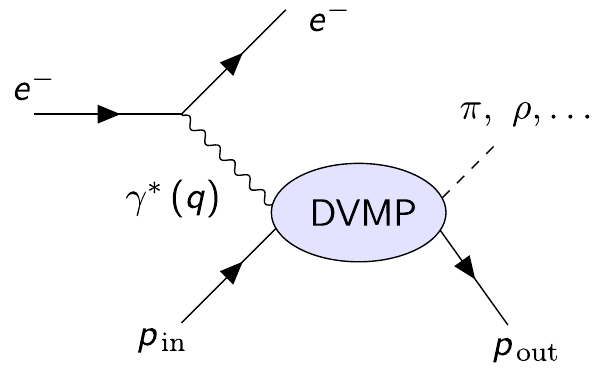}
  \caption{Deep Virtual Meson Production. This process is similar to DVCS but encompasses an additional difficulty as it also depends on the meson distribution amplitude.}
  \label{fig:DVMP}
\end{figure}

DVCS and TCS are the main processes whose probes and produced particles are described by quantum electrodynamics. They are therefore deeply studied since they do not involve complication coming from QCD bound states. If we relax this constraint, and allow ourselves to complicate the experimental processes by including additional bound states in the final state, we are naturally brought to consider Deep Virtual Meson Production (DVMP). This process is similar to DVCS, but replacing the outgoing photon by a outgoing meson, as illustrated on fig. \ref{fig:DVMP}. Thus, it also factorises\footnote{Factorisation has been proven for longitudinally polarised virtual photons only.} \cite{Collins:1996fb} into a hard part computable in perturbation theory (known today at NLO \cite{Mueller:2013caa}), a GPD, but also another non-perturbative correlator called the meson distribution amplitude (DA). This DA obviously complicates the extraction of GPDs from DVMP.

In principle, mesons act as a filter, allowing us to access different combinations of GPDs. It was originally thought that detecting a longitudinally polarised vector meson would allow us to extract singlet combinations of GPDs, a pseudo-scalar meson, the non-singlet ones, and a transversely polarised vector mesons, chiral-odd GPDs.
However, it was shown that transversally polarised mesons production vanishes at all-order of perturbation theory at leading twist \cite{Diehl:1998pd,Collins:1999un}.
This precludes any extraction of chiral-odd GPDs at leading twist with the sole production of vector mesons, but also provide a way to assess the potential impact of higher-twist contributions.
And the latter can be large as $\sigma_L$ and $\sigma_T$, the component of the cross-section due to respectively a longitudinally or transversally polarised virtual photon, are equal for $\rho_0$ production at low $Q^2 \simeq 1.5GeV^2$ (see \emph{e.g.} ref. \cite{Favart:2015umi}).
The ratio $\sigma_L/\sigma_T$ then increases with $Q^2$, albeit slower than expected \cite{Favart:2015umi}, highlighting the potential role of higher twist distributions. The exact origin of this mild $Q^2$ behaviour remains to be determined.

In the pseudo-scalar case, the situation is even worse. Indeed, the measurements performed at Jefferson Laboratory with a 6 GeV electron beam on pion electroproduction highlight the dominance of $\sigma_T$ over $\sigma_L$ \cite{Defurne:2016eiy}. And recent results from the 12 GeV beam upgrade suggest that this is still the case at higher values, up to $Q^2 = 8.4 \textrm{GeV}^2$ \cite{JeffersonLabHallA:2020dhq}. This situation highlights that leading twist dominance happens at different values of $Q^2$ depending on the considered meson. It also stresses that in the case of the pion, a leading twist interpretation remains out of reach for now. Thus, \emph{models} of reaction mechanisms have been developed, relying on higher-twist contributions and chiral-odd GPDs \cite{Goldstein:2013gra,Goloskokov:2009ia}. However, one should keep in mind that these models do not rely on factorisation theorems.

DVMP is thus difficult to analyse. But it presents nevertheless a major interest regarding the deconvolution problem. Indeed, contrary to DVCS, the shadow GPDs associated with DVMP are expected to be significantly different at NLO and beyond (and also meson-dependent due to the convolution with the DA).
To illustrate this, we give the equivalent of CFF for DVMP, called transition form factors $\mathcal{T}_V$ (TFF) for a vector meson $V$, are given as \cite{Mueller:2013caa}:
\begin{align}
  \label{eq:TFF}
  \mathcal{T}_{V}^q \propto \varphi_V(u) \underset{u}{\otimes} T^q\left(u,\frac{x}{\xi}\right) \underset{x}{\otimes} H(x,\xi),
\end{align}
where $\varphi_V$ is the meson distribution amplitudes. At leading order, $T^q$ is separable:
\begin{align}
  \label{eq:LOTFF}
  T^q\left(u,\frac{x}{\xi}\right) & = T_\varphi(u) \times T_H\left(\frac{x}{\xi} \right) \nonumber \\
                                  & \propto \alpha_s(\mu^2) \frac{1}{1-u} \frac{1}{\xi-x-i\epsilon },
\end{align}
and yields the same shadow GPDs then LO DVCS. However, at NLO, $T^q$ is not separable anymore \cite{Mueller:2013caa}, and therefore present a completely different functional form compared to the DVCS case. We thus expect that NLO DVCS shadow GPDs will not be NLO DVMP shadow GPDs.
Therefore, joined extractions of DVCS and DVMP will reduce the phase space allowed for shadow GPDs, and thus improving theoretical systematic uncertainties. Such a study would be of a major interest, but remains to be done at the time of writing.

Until now, we have highlighted the case of production of light mesons through electroproduction, where the hard scale is provided by the large virtuality of the incoming photon. But in the case of heavy mesons, such as $J/\Psi$ or $\Upsilon$, the mass of the meson itself can be seen as the hard scale allowing factorisation. This connects photoproduction data of $J/\Psi$ and $\Upsilon$ to GPDs. Formally speaking, factorisation has been proven only at NLO, and not at all orders \cite{Ivanov:2004vd}. Neglecting intrinsic heavy flavour distributions, these processes would lead to probe gluon GPDs at LO\footnote{Gluon GPDs also appear at LO in the description of light vector mesons such as $\rho^0$, but are mixed with the singlet combination of quarks.} at scales close to the charm or bottom threshold. Therefore, feasibility studies have been performed in order to measure these photoproduction reactions in UPC at the LHC \cite{Lansberg:2018fsy}. These studies yield optimistic results and have triggered enthusiasm within the hadron physics community \cite{Chapon:2020heu}.

Finally, let us mention processes with more than one particle produced in the final states have been suggested (such as the production of a pair of photon \cite{Pedrak:2020mfm,Grocholski:2021man,Grocholski:2022rqj} or a photon-meson pair \cite{Boussarie:2016qop,Duplancic:2018bum} in direct or reverse kinematics \cite{Qiu:2022bpq}). Processes with two hard scales such as Double DVCS \cite{Guidal:2002kt,Belitsky:2002tf,Pire:2011st} have also been suggested. However, these processes are more difficult to measure and feasibility studies at current and future facilities are ongoing.


%% file: section_perspectives.tex
Through these lecture notes, we have highlighted both the wealth and complexity of the GPD fields. After two decades of studies, GPDs remain mostly unconstrained. This is due to both the difficulties to measure exclusive processes and the necessity to develop advanced technical tools allowing one to handle multiple channels at the same time, with definite theoretical assumptions (twist expansion, $\alpha_s$ expansion, etc). In such a context, open-source integrated softwares dedicated to GPDs phenomenology, such as PARTONS\footnote{partons.cea.fr} \cite{Berthou:2015oaw} and GeParD\footnote{gepard.phy.hr} \cite{Kumericki:2016ehc} represent the best way to proceed, allowing everyone to take advantage of validated numerical tools.

Finally, let us mention that lattice-QCD simulations will probably play an important role in the forthcoming decade. Indeed, in the last decade, new techniques have emerged, allowing to match euclidean correlators with lightcone ones. The two most commonly explored techniques are the quasi-distribution formalism \cite{Ji:2013dva} and the pseudo-distribution one \cite{Radyushkin:2017cyf}. Both of them, in principle, unlock the possibility to assess GPDs using lattice-QCD simulations. Pioneering results on GPDs were obtained \cite{Alexandrou:2020zbe,Lin:2020rxa}, but remain at a too early stage to be exploited and compared with phenomenological data. Nevertheless, it is an important step forward that has been recognised by the community \cite{Constantinou:2020hdm} and may become an important source of knowledge on the multidimensional structure of the nucleon before the EIC starts running.
